\documentclass{article} % For LaTeX2e
\usepackage{iclr2025_conference,times}

\usepackage{amsmath,amsfonts,bm}

\def\eqref#1{equation~\ref{#1}}
\def\1{\bm{1}}

\DeclareMathAlphabet{\mathsfit}{\encodingdefault}{\sfdefault}{m}{sl}
\SetMathAlphabet{\mathsfit}{bold}{\encodingdefault}{\sfdefault}{bx}{n}

\usepackage[utf8]{inputenc} % allow utf-8 input
\usepackage[T1]{fontenc}    % use 8-bit T1 fonts
\usepackage{hyperref}       % hyperlinks
\usepackage{url}            % simple URL typesetting
\usepackage{booktabs}       % professional-quality tables
\usepackage{amsfonts}       % blackboard math symbols
\usepackage{nicefrac}       % compact symbols for 1/2, etc.
\usepackage{microtype}      % microtypography
\usepackage{xcolor}         % colors

\usepackage{amsmath}
\usepackage{amssymb}
\usepackage{mathtools}
\usepackage{amsthm}
\usepackage[section]{placeins}  % Float barrier
\usepackage{algorithm}
\usepackage{algorithmic}
\usepackage{multirow}

\newtheorem{theorem}{Theorem}
\newtheorem{corollary}{Corollary}[theorem]

\title{A mechanistically interpretable neural network for regulatory genomics}

\author{
    Alex M. Tseng\\
    \texttt{tseng.alex@gene.com}\\
    \And G\"okcen Eraslan\\
    \texttt{eraslan.gokcen@gene.com}\\
    \And Tommaso Biancalani\\
    \texttt{biancalani.tommaso@gene.com}\\
    \And Gabriele Scalia\\
    \texttt{scalia.gabriele@gene.com}\\
    \\
    Biology Research | AI Development\\
    Genentech
}

\iclrfinalcopy % Uncomment for camera-ready version, but NOT for submission.
\begin{document}

\maketitle

\begin{abstract}
Deep neural networks excel in mapping genomic DNA sequences to associated readouts (e.g., protein--DNA binding). Beyond prediction, the goal of these networks is to reveal to scientists the underlying motifs (and their syntax) which drive genome regulation. Traditional methods that extract motifs from convolutional filters suffer from the uninterpretable dispersion of information across filters and layers. Other methods which rely on importance scores can be unstable and unreliable. Instead, we designed a novel \textit{mechanistically interpretable} architecture for regulatory genomics, where motifs and their syntax are directly encoded and readable from the learned weights and activations. We provide theoretical and empirical evidence of our architecture's full expressivity, while still being highly interpretable. Through several experiments, we show that our architecture excels in \textit{de novo} motif discovery and motif instance calling, is robust to variable sequence contexts, and enables \textit{fully interpretable} generation of novel functional sequences.
\end{abstract}

\section{Introduction}
\label{intro}
Transcription factors (TFs) are proteins that regulate gene activity by recognizing and binding to specific short DNA sequence patterns---or ``motifs''---in the genome \citep{Lambert2018-gq,Siggers2014}. High-throughput experiments measure regulatory activity---such as protein--DNA binding or associated readouts---across the genome \citep{ENCODE_Project_Consortium2012-wb}. In general, the regulatory function of a DNA sequence is defined by the combination of motifs in that sequence. Understanding the motifs (and their syntax) which regulate the genome is thus crucial for many scientific and medical tasks, such as disease diagnosis and design of novel therapies (e.g., with CRISPR). Extracting motifs and syntax from these experiments, however, is challenging. Motifs and their configurations have a soft syntax (i.e., density, spacing, orientation, and co-binders), which induces a particular function such as TF binding, and these syntactical rules are dependent on surrounding context and cell type.

Thus, accurately predicting genome regulation from DNA sequence requires expressive models to capture the complexities of motifs and their syntax. In particular, deep neural networks (DNNs) have achieved state-of-the-art performance in mapping DNA sequences to TF binding and associated readouts \citep{Kelley2016,Avsec2021-yf,Linder2023}. These models take a DNA sequence as an input, and predict a label measured by a biological experiment (e.g., a binary label denoting if a TF bound to that sequence). The common goal of these genomic DNNs is to ultimately identify the fundamental code (i.e., motifs and their syntax) underpinning genome regulation \citep{Novakovsky2022r}. Notably, recent work has shown that motif discovery from these DNNs has far surpassed the ability of traditional statistical methods \citep{Tseng2022}.

There are two main classes of methods that extract motifs from a trained DNN. As these DNNs are almost universally convolution-based in early layers \citep{Novakovsky2022r}, a very common approach is to extract a motif from each first-layer filter \citep{Kelley2016}. This method, however, suffers from the critical limitation that information (including motifs) tends to be \textit{distributed} or \textit{dispersed} across filters and layers, thus there is no guarantee that any single filter will learn a biologically meaningful motif (Figure \ref{fig:motif-disc}a). The second class of methods relies on importance scores, which attempt to measure the contribution of individual DNA bases to the output prediction, with the assumption that bases in motifs have elevated importance. By integrating throughout the whole DNN, importance scores bypass the problem of distributed information. Unfortunately, importance scores are known to be highly unstable and unreliable approximations of a model's decision making \citep{Ghorbani2017,Alvarez-Melis2018}, and in genomic DNNs, importance scores typically only show noisy and fragile motif instances (Supplementary Figure \ref{suppfig:impscores}) \citep{Tseng2020a}. In practice, to improve robustness, motif discovery via importance scores requires complex pipelines composed of many computationally expensive steps, which tend to be delicate and require constant human intervention \citep{Novakovsky2022r,Shrikumar2018}.

In this work, we propose a new method of recovering motifs from a DNN, based on \emph{mechanistic interpretability}. Mechanistic interpretability (MI) has recently emerged as a key research direction to explain complex models \citep{Bereska2024}. Our method, \textbf{Analysis of Regulatory Genomics via Mechanistically Interpretable Neural Networks (ARGMINN)}, enables motifs \textit{and} their syntax to be directly readable from the network architecture, without compromising expressivity or relying on complex \textit{post hoc} pipelines. In Section \ref{sec:arch}, we formally describe the ARGMINN architecture, including several novel architectural contributions including: \textbf{1)} a regularizer designed to ensure that the first layer's convolutional-filter weights directly encode a non-redundant set of relevant motifs; and \textbf{2)} a modified attention mechanism which reveals motif instances and their syntax in any query sequence with a single forward pass.

In Section \ref{sec:results}, we show experimental results which demonstrate ARGMINN's interpretability and its main contributions to regulatory genomics, including: \textbf{1)} superior motif discovery \textit{and} motif-instance/syntax analysis compared to existing approaches; \textbf{2)} robustness against natural or adversarial sequence modifications; and \textbf{3)} the novel ability to perform \textit{fully interpretable} sequence generation. In Section \ref{sec:theory}, we provide theoretical results on ARGMINN's expressivity, showing that it can learn any motifs and syntax, whereas previous MI architectures cannot.

\section{Related Work}
\label{sec:related}

Practically all prevalent genomic DNN architectures have convolutional filters as the first layer \citep{Kelley2016,Avsec2021-yf,Linder2023}. To recover motifs, most works directly visualize the filter weights or average subsequences which highly activate each filter \citep{Alipanahi2015,Kelley2016}. Although this has shown some limited success, these methods assume that each filter learns one motif, and each motif is learned by one filter. This is generally not true because---without special constraints---motifs are typically learned in a \textit{distributed} fashion, where each motif is learned across many filters and layers (Figure \ref{fig:motif-disc}a).

As a result, more sophisticated \textit{post hoc} pipelines were developed to extract motifs from trained DNNs. These pipelines integrate over the entire DNN to compute importance scores across the dataset (e.g., via integrated gradients or DeepLIFTShap \citep{Sundararajan2017,Shrikumar2017}), resulting in an importance score at each position for each sequence. These scores are then segmented into high-importance regions as putative motif instances. Due to the noisiness of importance scores, however, these instances must be clustered into clean motifs by tools like MoDISco \citep{Shrikumar2018}. Each step of this pipeline is computationally expensive, and for most datasets, the time required to recover motifs is typically over an order of magnitude longer than the time needed to train the model. Furthermore, these pipelines heavily rely on importance scores from a black-box model, which can be extremely fragile, as importance scores frequently fail to reveal a model's true decision-making process \citep{Ghorbani2017,Kindermans2017,Alvarez-Melis2018,Tseng2020a}.

% To perform motif instance calling (i.e., identifying the location and identity of motifs within a sequence), current methods first perform motif discovery, and then instance-calling tools such as FIMO \citep{Bailey2015} scan query sequences for the discovered motifs.

Within the field of explainable AI, there has been some burgeoning work exploring MI, where the patterns and rules learned by a DNN are reflected in its physical computation \citep{Bereska2024}. In particular, our work is a type of \textit{intrinsic} MI, where the DNN's architecture (e.g., weights and activations) directly encode learning \citep{Liu2023,Barbiero2023,Kasioumis2021}. This is typically done by increasing sparsity, modularity, and the proportion of monosemantic neurons (i.e., neurons which learn a single concept). DNNs may also improve their intrinsic MI by learning logical rules in a more structured way \citep{Riegel2020,Friedman2023}.

Despite these promising works, constructing intrinsically MI architectures for very general problems remains difficult. However, focusing on more constrained predictive tasks (e.g., motif discovery) makes MI more feasible, allowing us to restrict computation and the solution space. ARGMINN uses similar principles as other intrinsic MI works (e.g., sparsity, modularity, monosemanticity, and explicit logic), but in a biologically grounded manner, making it both highly interpretable and expressive.

Recently, the ExplaiNN architecture also aimed to interpretably learn motifs from DNA sequences \citep{Novakovsky2022}. ExplaiNN consists of a set of single-filter convolutional networks. Each network ideally learns a distinct motif, and outputs a scalar summarizing binding strength over the input. The final output is a learned linear combination of these scalars. In later sections, we will show that ARGMINN surpasses ExplaiNN in both interpretability and expressivity.

Our work also relates to concept bottleneck models (CBMs) \citep{Koh2020,Ghorbani2019,Kim2017}, which force decisions to be based on interpretable concepts from an intermediate layer. While CBMs can provide meaningful explanations, they require pre-defined concepts and concept-labeled inputs, which are labor-intensive to obtain. In our work, ARGMINN can be viewed as a type of CBM, but the concepts are motifs, and \textit{the concepts are learned entirely from the data}, thus overcoming the typical limitations of CBMs. Furthermore, ARGMINN elucidates syntax between motifs---rules relating to \textit{positioning} between concepts---in addition to the motifs themselves. 

\begin{figure}[t]
\centering
\includegraphics[width=\columnwidth]{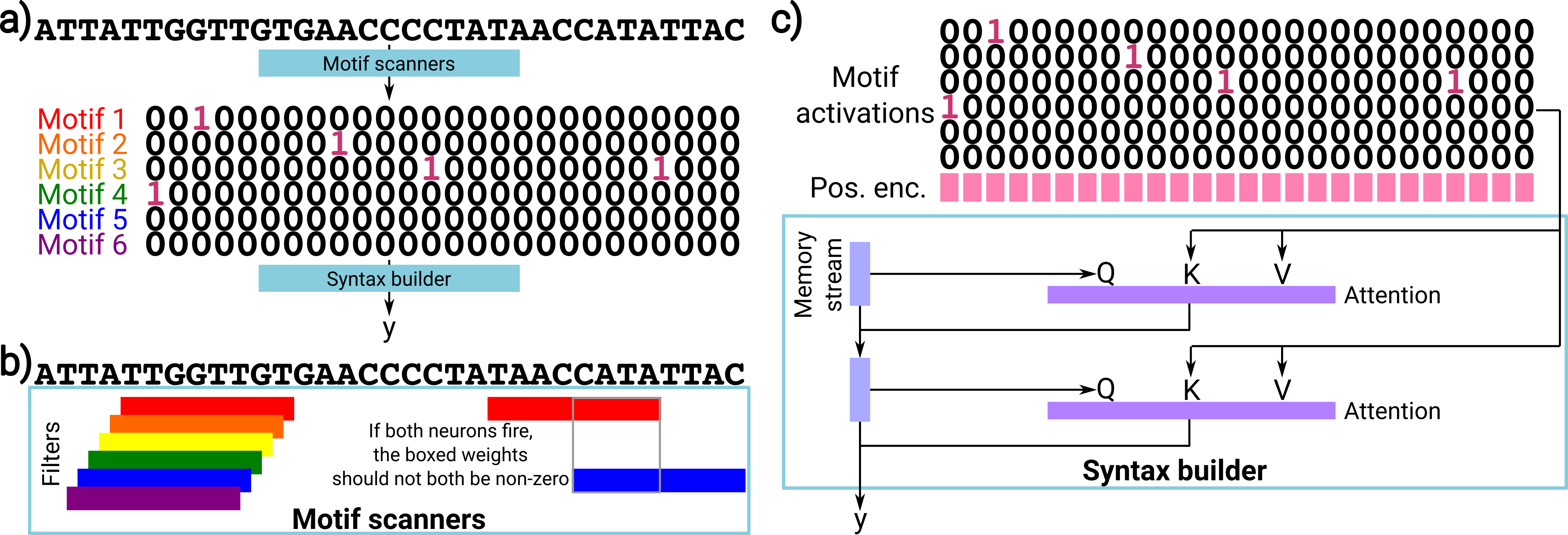}
\caption{\small Schematic of the ARGMINN architecture. \textbf{a)} The motif-scanner module produces activations denoting which motifs were found at each position, where activation magnitude reflects match strength. The activations are passed to the syntax builder, which learns higher-order logic between motif instances for the final prediction. \textbf{b)} The motif-scanner module is a single convolutional layer which learns all motifs \textit{de novo}. Regularization ensures that each filter learns one motif (and \textit{vice versa}), penalizing different filters from activating based on the same underlying subsequences. \textbf{c)} The syntax builder is a series of uniquely designed attention layers. An explicit memory stream tracks the model's state. Each attention layer derives a single query vector from the memory stream, and key/value vectors from the original activations, to update the memory stream.}
\label{fig:arch}
\end{figure}
\section{ARGMINN Architecture}
\label{sec:arch}

We propose a mechanistically interpretable DNN architecture designed to 1) accurately predict regulatory function (e.g., protein binding) from DNA sequence; 2) reveal crucial motifs responsible for function across the dataset; and 3) reveal motif instances and their syntax in any given query sequence. Importantly, 2) and 3) are directly encoded in the model's weights and activations.

Our architecture consists of two modules trained end to end (Figure \ref{fig:arch}a). The \textit{motif-scanner} module is a single convolutional layer which identifies motifs from the input sequence. We developed a novel regularizer for the filters so that the filter weights directly encode non-redundant motifs, thereby accomplishing goal 2) above. The convolutional activations are passed to the second module---the \textit{syntax builder}---consisting of several layers of modified attention that assemble the syntax and logic between motifs to produce a final prediction. The syntax builder is designed such that the activations and attention scores immediately reveal motif instances and syntax, thus achieving goal 3).

\subsection{Module 1: Motif Scanners}

The ``motif scanners'' are a set of $n_{f}$ convolutional filters of width $w$, which are scanned across the input sequence to produce a set of activations (Figure \ref{fig:arch}b).  Thus, a filter's activation is maximized when it scans over a one-hot-encoded sequence that exactly matches the encoded motif (motifs are learned \textit{de novo}). The activations are then thresholded by a ReLU function. The additive bias parameter of the filters and the ReLU allow the network to selectively cut off weak matches, thereby producing more sparse activations. For a 1-hot encoded DNA sequence $S\in\{0,1\}^{\ell\times 4}$, the convolutional weights $W\in\mathbb{R}^{n_{f}\times w \times 4}$ (and bias $b\in\mathbb{R}^{n_{f}}$) yield activations $A\in\mathbb{R}_{+}^{(\ell-w+1)\times n_{f}}$:

\begin{equation}
    A = \text{ReLU}(\text{Conv}(S, W, b)).
\end{equation}

Importantly, the filters are regularized so that each filter learns one distinct motif, and each motif is learned by one filter. This allows motifs to be directly read from the filter weights after training. We propose a novel secondary objective which penalizes different filters from activating on overlapping sequences. This is combined with a simple L1-penalty on the filter weights themselves to induce sparse filters that directly reveal \textit{distinct, non-redundant} motifs.

At each position $i$ in sequence $S$, each filter aggregates values $S[i,i+w]$. Let $a = \text{argmax}\{A_{i}\} \in \{0,\ldots,n_{f}-1\}$ be the index of the maximally activated filter at $i$. For all other filters $b\neq a$, if filters $a$ and $b$ both achieve non-zero activation in the same neighborhood of $S$, then they should not both have non-zero weights in the overlapping region (i.e., they should not be attending to the same part of the sequence). In other words, every position of $S$ is ``protected'': at most one filter can activate while attending to any position. If another filter is activated nearby, then its weights should not be attending to the protected part of the sequence. Our \emph{filter-overlap} regularization then can be defined as the following loss function:

\begin{multline}
    \label{eq:filterreg}
    \mathcal{L}_{o}(A, W) = \sum\limits_{i=0}^{\ell - w}\sum\limits_{b \neq \text{argmax}\{A_{i}\}}\sum\limits_{j=i-(w-1)}^{i+(w-1)}\Big[A_{j, b}\\
    \Vert W_{\text{argmax}\{A_{i}\}}[\max\{0, j - i\}, w - 1 - \max\{0, i - j\}]\Vert_{1}\\
    \Vert W_{b}[\max\{0, i - j\}, w - 1 - \max\{0, j - i\}]\Vert_{1}\Big].
\end{multline}

At each position $i$, we compare the filter weights of the maximally activated filter $W_{a}$ with the weights of all other filters ($W_{b}$), at every possible overlapping window $j$. We penalize the L1 norm of the overlapping weights, multiplied by the activation of filter $b$ (i.e., if $W_{b}$ is not activated, there is no penalty). Importantly, this is a \textit{soft} regularization which the model can choose to ignore if needed for performance. This regularization helps prevent: 1) two filters learning the same motif (or the same part of the same motif); and 2) two filters learning a motif in an interleaved fashion. However, our regularization still allows for a long motif to be learned by two filters, split somewhere down the middle (or similarly, two half-sites directly next to each other, each learned by one filter).

In practice, this loss is computed efficiently by caching all possible windows of weight sums (for each value of $j$) once, and at each $i$ scaling the window products with the activation of $W_{b}$.

\subsection{Module 2: Syntax Builder}

After the motif-scanner module, positional encodings $P$ are concatenated to the activations $A$. The second module of ARGMINN is the syntax builder, which consists of $n_{L}$ layers of a custom attention mechanism that learns syntax between motifs (Figure \ref{fig:arch}c). In contrast to typical attention, our modified attention layer has an explicit ``memory stream'' $m_{l}$ which is updated after each layer (inspired by \citet{Friedman2023}). Each layer derives a \textit{single} query from $m_{l}$, resulting in a linear vector of attention scores rather than a quadratic matrix (this improves both interpretability and efficiency). Importantly, every layer derives key and value vectors \textit{directly} from the original ``tokens'' (i.e., $A\Vert P$). Each attention layer can be described as follows:

\begin{equation}
\begin{split}
    \label{eq:att}
    q_{l} := W_{Q,l}m_{l-1},\quad K_{l} := W_{K,l}[A \Vert P], \quad V_{l} := W_{V,l}[A \Vert P]\\
    a_{l} := \frac{K_{l}q_{l}}{\sqrt{d_{q_{l}}}}, \quad m_{l} := m_{l-1} + \text{MLP}(a_{l}V_{l}),
\end{split}
\end{equation}
where $m_{0}$ is initialized as a vector of ones, and $d_{q_{l}}$ is the dimension of the query vector. We also include $n_{h}$ attention heads, but do not show the reshaping operations above for clarity.

Each layer can attend to multiple motifs (due to the multiple attention heads), and successive layers allow the model to capture \textit{interactions} between motifs (e.g., with $k$ layers, the model can reason about $k$th-order interactions between motifs). All together, our final loss function becomes:

\begin{equation}
    \label{eq:finalloss}
    \mathcal{L}(S, A, W) = \mathcal{L}_{pred}(f(S), y) + \lambda_{o}\mathcal{L}_{o}(A, W) + \lambda_{l}\Vert W\Vert_{1}.
\end{equation}

\section{Experimental Results}
\label{sec:results}

In this section, we show experimental results demonstrating ARGMINN's advantages in motif discovery, motif instance calling, robustness, and interpretable generation.

\subsection{Improved motif discovery}

To extract motifs from ARGMINN, we applied the procedure from \citet{Kelley2016} to the first-layer convolutional filters. Specifically, we obtained a motif from each filter by averaging test-set subsequences which highly activate the filter (filters which were never activated in the test set were dropped) (Supplementary Methods \ref{app:supp-methods:analyses}).

\begin{figure}[h!]
\centering
\includegraphics[width=0.92\columnwidth]{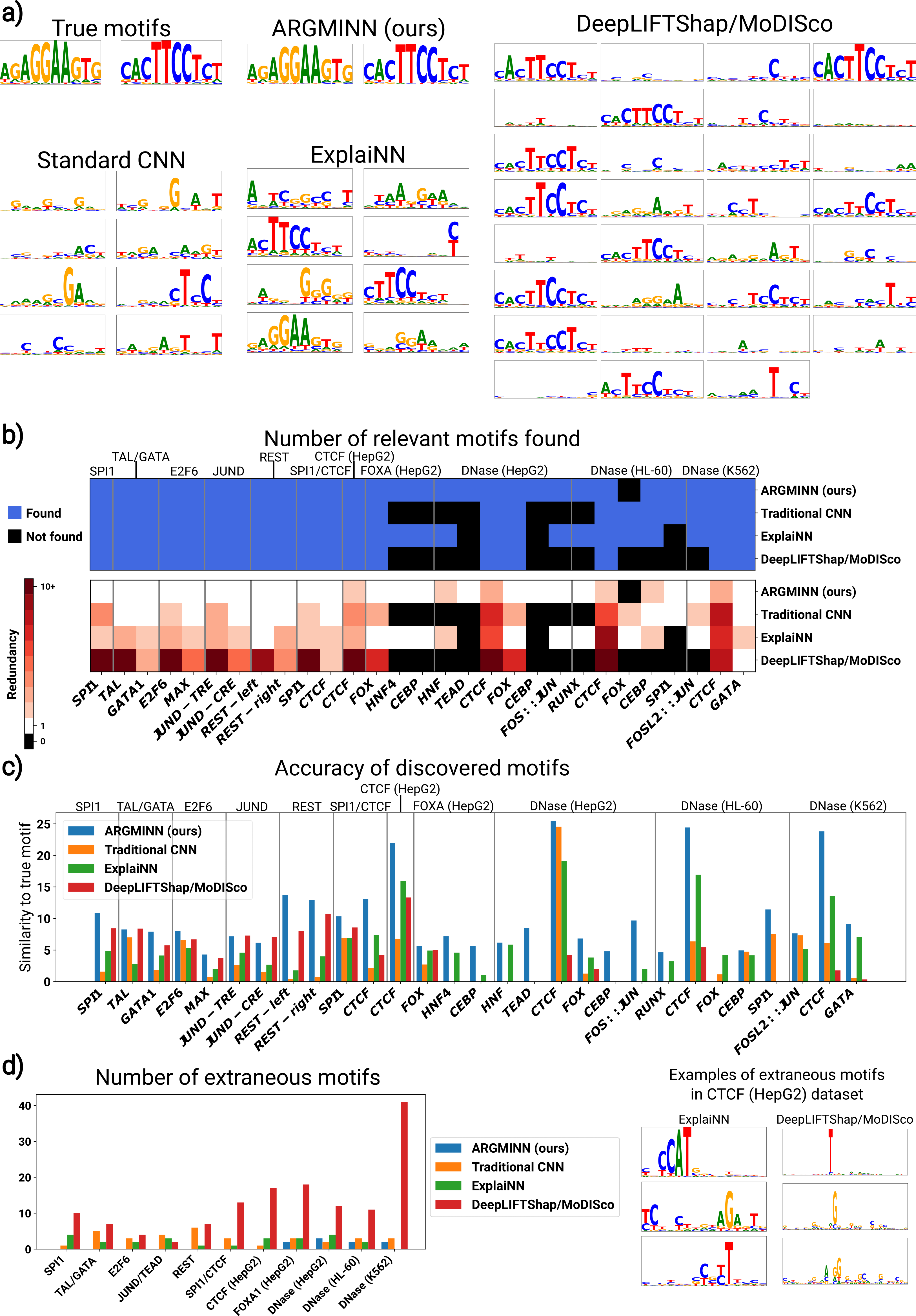}
\caption{\small Motif discovery. \textbf{a)} Example of SPI1 motifs discovered by ARGMINN, compared to interpreting the first-layer filters of a standard CNN, using ExplaiNN, and by clustering DeepLIFTShap importance scores using MoDISco. Note that MoDISco combines forward and reverse-complement orientations. \textbf{b)} For each dataset, we show whether or not each method successfully recovered each relevant motif (above), and the amount of redundancy as the number of times each motif was discovered (below). \textbf{c)} To quantify accuracy of the discovered motifs, for each relevant motif we show the maximum similarity to motifs discovered by each method. \textbf{d)} For each dataset, we show the number of extraneous motifs---those which do not match any known relevant motif---that each method discovered (left). We show a few examples of such extraneous motifs discovered for the CTCF (HepG2) experimental dataset (right).}
\label{fig:motif-disc}
\end{figure}

Over several simulated and real-world experimental datasets (Supplementary Methods \ref{app:supp-methods:data}), we compared the motifs discovered by ARGMINN to those identified by several other methods: interpreting the filters of a standard CNN, ExplaiNN \citep{Novakovsky2022}, and importance-score clustering via DeepLIFTShap and MoDISco \citep{Shrikumar2018}. We systematically matched each discovered motif to the closest known relevant motif. For simulated datasets, we matched to ground-truth motifs; for experimental datasets, we matched to the closest relevant human motif. ARGMINN identified \textit{known, biologically relevant motifs}, and compared to other methods, it generally missed the fewest relevant motifs and discovered the fewest redundant motifs (Figure \ref{fig:motif-disc}a--b, Supplementary Figures \ref{suppfig:more-disc-motifs-1}--\ref{suppfig:more-disc-motifs-4}, Supplementary Table \ref{supptab:motif-counts}). For example, ARGMINN trained on FOXA2 in HepG2 (a pioneer factor) revealed factors in the FOX, HNF4, and CEBP families, all known to co-localize or co-bind with FOXA \citep{Seachrist2021,Geusz2021,Liu2020}. Notably, in the experimental datasets, the singular most similar motif (an extremely strict requirement) to an ARGMINN filter is typically a known relevant motif (e.g., on this FOXA dataset, ARGMINN identified 4 motifs whose top match was a relevant motif, whereas the traditional CNN only identified 2, with much weaker similarities). Additionally, ARGMINN's motifs were generally most similar to the ground truth (Figure \ref{fig:motif-disc}c, Supplementary Table \ref{supptab:motif-sims}), thus highlighting their \textit{quality}. ARGMINN was also the method which consistently identified the fewest extraneous motifs---patterns which do not match any biologically relevant motif (Figure \ref{fig:motif-disc}d, Supplementary Table \ref{supptab:extra-motifs}).

In contrast, other methods significantly underperformed compared to ARGMINN. Even when the standard CNN or ExplaiNN encoded motifs in their filters, ARGMINN's filters were far more similar to the true motifs. ARGMINN also outperformed MoDISco \citep{Shrikumar2018}, which---despite recovering many relevant motifs---consistently identified less-accurate motifs than ARGMINN (Figure \ref{fig:motif-disc}c). Due to the unreliability and noisiness of the importance scores themselves, as well as the frailty of clustering, MoDISco also found \textit{many} redundancies and extraneous motifs (Figure \ref{fig:motif-disc}b, \ref{fig:motif-disc}d).

\subsection{Improved motif instance calling and syntax discovery}

\begin{figure}[h]
\centering
\includegraphics[width=\columnwidth]{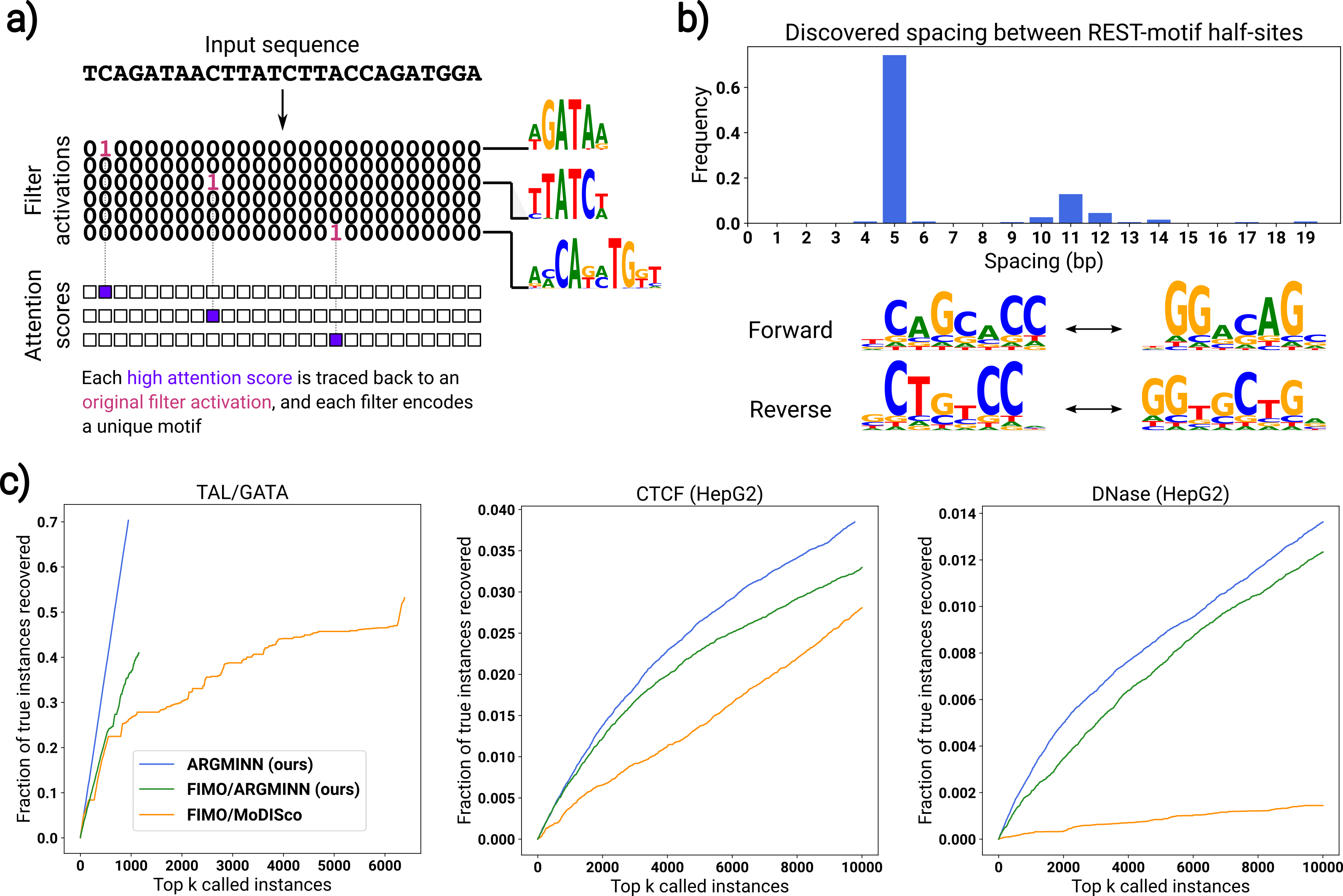}
\caption{\small Motif instance calling and syntax discovery. \textbf{a)} ARGMINN calls motif instances in any query sequence in only a forward pass. High attention scores in any attention layer trace directly back to original filter activations, which directly map to sequence motifs. \textbf{b)} After training on an experimental dataset of REST binding in HepG2, ARGMINN revealed the unique binding syntax of REST in both the forward and reverse-complement orientations, where the half-sites (left and right) bind either adjacently or around 9--14 bp apart. \textbf{c)} We compare motif instances discovered by ARGMINN to the traditional approach of using MoDISco to discover motifs and subsequently scanning for them with FIMO. We rank motif instances by confidence (attention score from ARGMINN, or FIMO hit q-value), and compute the fraction of true instances that are covered in a top-$k$ fashion. We also compare to motif instances discovered by scanning for ARGMINN-discovered motifs with FIMO.}
\label{fig:motif-hits}
\end{figure}

With previous DNN-based methods, identifying motif syntax required first learning motifs (e.g., via MoDISco) and then scanning sequences to ``call'' motif instances. Not only is this computationally expensive, but instance calling by sequence scanning tends to be highly inaccurate (e.g., due to partial hits or missing context which the DNN would have considered).

Instead, ARGMINN reveals motif instances by tracing attention scores from a single forward pass on any sequence (Figure \ref{fig:motif-hits}a). Since every attention layer derives keys/values directly from the motif-scanner activations, high attention scores directly point to the precise motif instances which the network deemed important for prediction. Specifically, for any input query sequence, we examine the attention scores across all layers/heads on the forward pass. For each high score (e.g., $> 0.9$), we identify the corresponding sequence position. We then check the motif activations at that position, and call a motif instance if a motif/filter has high activation (Supplementary Methods \ref{app:supp-methods:analyses}). For example, trained on an experimental dataset of REST binding, ARGMINN directly recovered the unique binding syntax---including spacing preferences---of the two halves of the REST motif in both directions/orientations \citep{Tang2021} (Figure~\ref{fig:motif-hits}b).

We then quantitatively evaluated motif instances identified by ARGMINN versus a traditional pipeline. Namely, using FIMO \citep{Bailey2015}, we called instances of MoDISco-discovered motifs \citep{Shrikumar2018} in test sequences. For each method, we ranked instances by confidence and computed the number of ground-truth instances recovered by the top-$k$ called instances. For experimental datasets, we used independently derived binding footprints as ground truth \citep{Vierstra2020}. In general, ARGMINN's motif instances were far more accurate than those found by the baseline (Figure \ref{fig:motif-hits}c, Supplementary Figure \ref{suppfig:more-motif-hits}, Supplementary Table \ref{supptab:motif-hit-pr}). To gain further intuition, we also compared to motif instances found by using FIMO to scan for ARGMINN-discovered motifs. ARGMINN identified more accurate motif instances than FIMO, even when FIMO was given the same set of ARGMINN motifs. This comparison also shows the direct benefit of using ARGMINN to perform motif instance calling, compared to the traditional method of sequence scanning.

Finally, to further demonstrate that ARGMINN's attention scores match underlying biological signal, we show that the positions of high attention scores in experimental test-set sequences closely track the measured biological strength of protein binding along the sequence (Supplementary Figure \ref{suppfig:att-score-peaks}).

\subsection{QTL prioritization}

\begin{figure}[h]
\centering
\includegraphics[width=0.8\columnwidth]{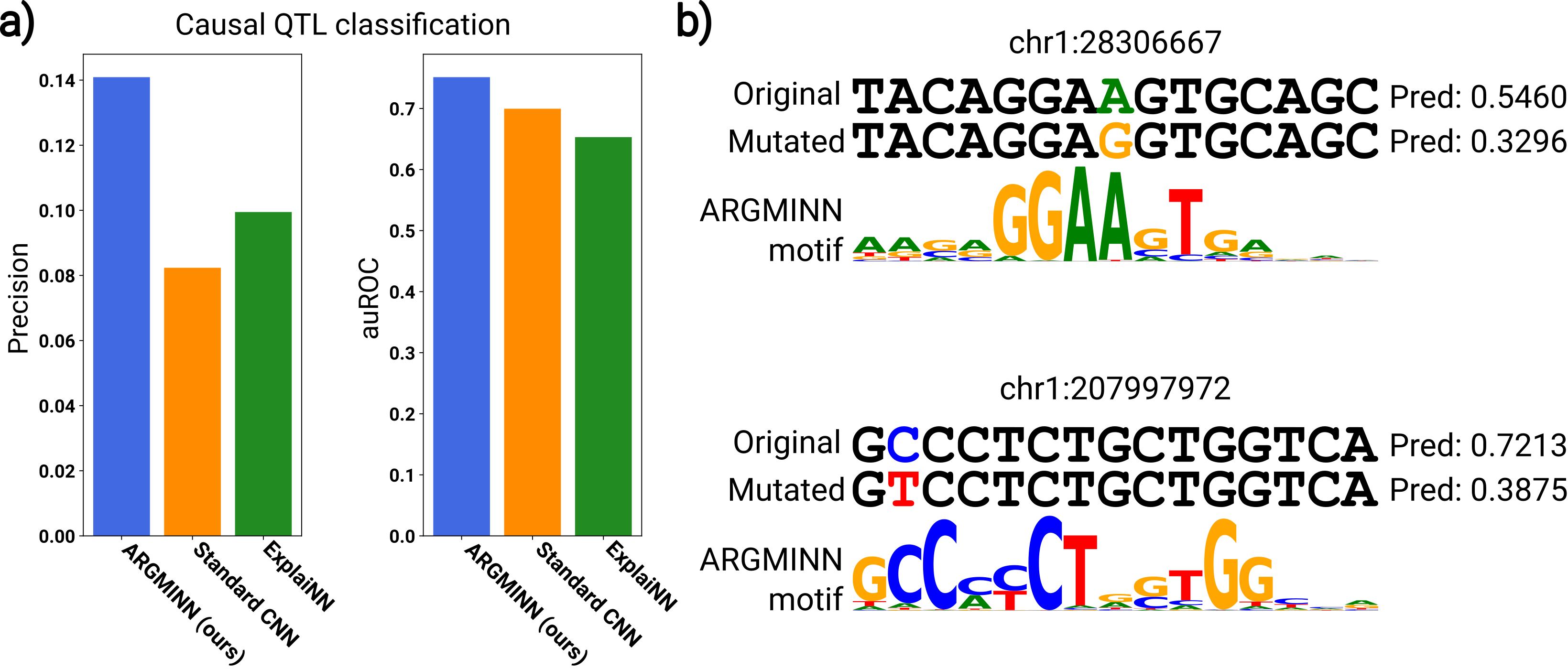}
\caption{\small QTL classification. \textbf{a)} On a set of known DNase-sensitivity QTLs, we evaluated the ability of ARGMINN to prioritize true causal dsQTLs by quantifying the difference in predictions with and without the dsQTL mutation. \textbf{b)} We show two examples of causal dsQTLs, which fall in ARGMINN-discovered motif instances. In both cases, by making the dsQTL mutation, the ARGMINN-discovered binding site is disrupted, leading to a reduced prediction of accessibility.}
\label{fig:qtls}
\end{figure}

Next, we evaluated ARGMINN's ability to classify and prioritize a set of causal DNase-sensitivity quantitative trait loci (dsQTLs) from a background set of non-causal dsQTLs. dsQTLs are mutations which change the DNase accessibility of a sequence. A predictive model which makes decisions based on biological signals (and not spurious correlates) should predict a larger change for causal dsQTLs compared to non-causal dsQTLs, which are merely correlated with the causal changes. On a held-out chromosome, we found that compared to other architectures, ARGMINN was capable of classifying/prioritizing the causal dsQTLs much more accurately (Figure \ref{fig:qtls}a). In particular, these causal dsQTLs tended to overlap specific motif instances which ARGMINN used for its prediction of accessibility (Figure \ref{fig:qtls}b). This further shows that ARGMINN makes biologically meaningful predictions on sequence changes, based on interpretable motif biology rather than spurious signals.

\subsection{Robustness of ARGMINN}

Because ARGMINN makes decisions based on biologically meaningful motifs, it is more robust to background variations. On several simulated datasets, we trained ARGMINN (along with a standard CNN and ExplaiNN) with a 50\% GC background. We then computed predictive performance on test sequences with GC content ranging from 5\% to 95\%, thus simulating natural variation in background composition. ARGMINN's performance suffered the least, with a higher and tighter distribution of performance in general (Supplementary Figure \ref{suppfig:robust}a).

Furthermore, because traditional CNNs learn filters which do not accurately represent motifs, they are prone to adversarial attacks (Supplementary Methods \ref{app:supp-methods:analyses}). On our SPI1 dataset, we trained a CNN to achieve \textbf{88\%} accuracy. We then easily constructed many sequences containing short substrings which highly activate its filters, but without any instances of the SPI1 motif. On this set of sequences, the standard CNN's accuracy dropped to \textbf{55\%}. Reflexively, we also easily designed sequences containing the SPI1 motif, but we inserted substrings into the background which strongly deactivate the filters, leading the CNN's accuracy to drop to \textbf{48\%}. In contrast, because ARGMINN encodes meaningful biological motifs in each filter, it remains robust against such an attack (i.e., one cannot easily identify highly-activating or anti-activating sequences which trick the model into giving the wrong prediction).

Finally, we used Ledidi \citep{Schreiber2020} to generate sequences which would bind to SPI1. Ledidi is a gradient-based sequence-design method, which optimizes input sequences to maximize the probability of a positive prediction from a given model. Applying Ledidi on the ARGMINN architecture generated sequences with the strongest instances of the true SPI1 motif (Supplementary Figure \ref{suppfig:robust}b). The true binding strength of ARGMINN's Ledidi-generated instances was significantly higher than those generated from the standard CNN (\mbox{$p = 8.95\times 10^{-18}$}) and ExplaiNN (\mbox{$p = 5.53\times 10^{-146}$}). This is likely due to the fact that ARGMINN is explicitly trained to focus on the relevant motifs, whereas the other models often focus on spurious, non-motif sequence patterns, resulting in ARGMINN's being a more robust oracle for optimization.

\subsection{Interpretable design of novel functional sequences}

\begin{figure}[h]
\centering
\includegraphics[width=0.8\columnwidth]{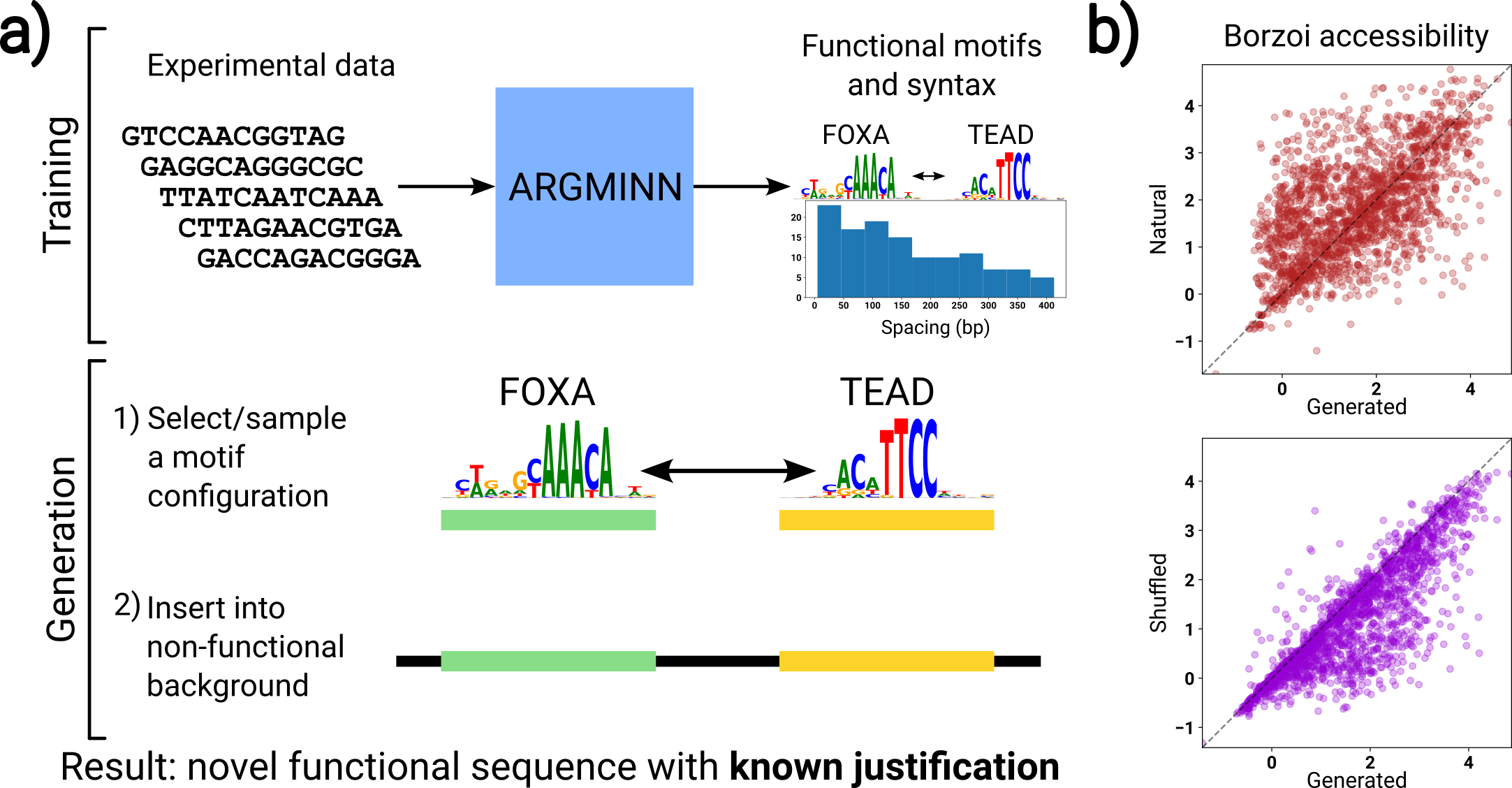}
\caption{\small Interpretable sequence design. \textbf{a)} After training on experimental data (e.g., HepG2 accessibility), ARGMINN reveals the functional motifs and their binding syntax which induce function. To generate a novel functional sequence, we insert motifs into a non-functional background, following the syntactical rules learned by ARGMINN. In this novel sequence, the mechanistic justification is fully controlled and known. \textbf{b)} We interpretably constructed novel HepG2-accessible sequences using ARGMINN and validated their accessibility using Borzoi. We compared the accessibility between generated sequences and natural sequences from the experiment (top), as well as between generated sequences and shuffled backgrounds (bottom).}
\label{fig:seq-design}
\end{figure}

Upon training, ARGMINN reveals the functional motifs and the syntatical rules for combining them which produce a prediction of binding or accessibility. As such, ARGMINN can be used for \emph{interpretable generation of novel functional sequences}. In contrast with sequence generation using traditional methods (e.g., gradient-based tools like Ledidi, or directed evolution) or non-interpretable generative models (e.g., diffusion models or autoencoders), we may start with ARGMINN interpretations and sample motif configurations to insert into non-functional backgrounds. This yields a generated sequence for which we have a complete understanding on the \textit{reason} why it is biologically active (e.g., it is active because it has the FOXA motif and TEAD motif 50 bp apart) (Figure \ref{fig:seq-design}a).

To demonstrate this ability, we trained ARGMINN to predict DNA accessibility in the HepG2 cell type from experimental data. On the test set, we then extracted the functional motifs and the syntactical rules and grammars between them. We sampled motifs and their configurations entirely from ARGMINN's discovered biology, and inserted them into non-functional sequence backgrounds (Supplementary Methods \ref{app:supp-methods:analyses}). Using the same model as a predictor of accessibility, we found that the newly generated sequences were predicted to be significantly more accessible than the shuffled non-functional backgrounds (\mbox{$p = 1.78\times 10^{-32}$}). Furthermore, these generated sequences were predicted to be \textit{just} as accessible as natural sequences identified to be highly accessible by the original biological experiment (\mbox{$p = 0.75$}) (Supplementary Figure \ref{suppfig:seq-design-val}c). We then repeated our \textit{in silico} validation with a completely independently trained model, Borzoi \citep{Linder2023}. Using the output head of Borzoi which predicts DNA accessibility in HepG2, we passed the same sequences through Borzoi to predict accessibility and confirmed the same trends: the generated sequences were far more accessible than the non-functional backgrounds, and were similar in accessibility to natural sequences identified by the experiment (Figure \ref{fig:seq-design}b). We found that this trend was also upheld by two additional independent oracles (Supplementary Figure \ref{suppfig:seq-design-val}a--b).

\section{Theoretical results}
\label{sec:theory}

For a MI architecture, ARGMINN is highly interpretable, yet uniquely retains high expressivity. Here, we show that ARGMINN is capable of learning any possible configuration of motifs---including positional and syntactic constraints---as long as the constraints are definable in first-order logic.

\begin{theorem}
\label{thm:argminn-express}
Every configuration of motifs/subsequences which is definable by a sentence in first-order logic (with positional variables) is recognizable by an ARGMINN classifier.
\end{theorem}

Furthermore, ARGMINN is more expressive than previous interpretable architectures for regulatory genomics, such as ExplaiNN \citep{Novakovsky2022}, which learns an output label from a linear combination of motif strengths:

\begin{corollary}
\label{corr:explainn-express}
There exists a configuration of first-order-logic-expressable motifs/subsequences which is \textnormal{not} recognizable by any ExplaiNN classifier.
\end{corollary}

We formally prove Theorem \ref{thm:argminn-express} and Corollary \ref{corr:explainn-express} in Appendix \ref{app:supp-proofs}. Here, we briefly sketch the proof of Theorem \ref{thm:argminn-express}. Intuitively, ARGMINN's motif-scanner module outputs motif binding strengths by encoding each motif's position-weight matrix \citep{Benos2002} in a different filter. With expressive positional encodings, the syntax-builder module learns syntax/interactions between motif instances. Each attention head learns one syntactical ``rule'' (i.e., possible motif combinations), which is built up in complexity over layers (i.e., combinations of $k$ motifs are learned by the $k$th layer). 

\subsection{Experimental follow-up to theoretical results}

To empirically reinforce our theoretical results, we consider the REST binding motif, which consists of two halves which must bind together (with variable spacings) (Figure \ref{fig:motif-hits}b). Notably, both halves need to bind in the same orientation and order (half sites cannot be mixed/matched). On a simulated REST dataset which explicitly tests these complex requirements, ARGMINN achieved \textbf{90.9\%} test-set accuracy, whereas ExplaiNN only achieved \textbf{73.4\%}. This demonstrates that ARGMINN is sufficiently expressive to capture complex grammars and syntax between motifs, which previous architectures could not.

\section{Discussion}
\label{discussion}

We illustrated ARGMINN's unique ability to reveal genome-regulatory motifs and their syntax directly from its weights and activations---an advantage which is entirely absent from traditional (non-MI) DNNs. We then compared the predictive performance of ARGMINN to standard CNNs and ExplaiNN (Supplementary Table \ref{supptab:perf}). It is generally well known that more interpretable models may suffer slightly in predictive performance \citep{Dosilovic2018,BarredoArrieta2020}. This is expected, as these models tend to base decisions on human-interpretable concepts (e.g., crucial motifs), instead of spurious signals which can be informative, but are not useful for understanding regulatory genomics (e.g., GC content, exceptionally rare motifs, etc.). Our empirical results fall in line with these expectations. On simulated datasets where we ensured that the only informative signals are motifs, ARGMINN achieved superior performance compared to the baselines. On experimental datasets with many more spurious signals, ARGMINN achieved competitive performance, but did not outperform its non-MI counterparts.

Additionally, to further show the benefit of our filter regularizer, we applied our regularization to a standard CNN. As a result, the CNN's filters also encoded relevant and non-redundant motifs (Supplementary Figure \ref{suppfig:cnn-filterreg}). This shows that \emph{our filter-overlap regularizer can be readily applied to standard genomic DNNs} to achieve more interpretable first-layer filters in general. Importantly, however, without the other architectural novelties of ARGMINN (Equation \ref{eq:att}), such a DNN would still not reap benefits such as motif-instance and syntax discovery.

Finally, we explored the robustness of ARGMINN to the loss weights (which are hyperparameters) for our regularizers. We found that over many orders of magnitude, ARGMINN's \emph{predictive performance and interpretability both remained robust to the loss weights} (Supplementary Figure \ref{suppfig:loss-weight-robust}). This is partially due to the design of these regularizers, which are aimed to \textit{synergize} with predictive performance and the learning of biological motifs, rather than compete with other losses.

Mechanistic interpretability is still a nascent field, and our research pioneers an architecture that enables direct interpretation with minimal \textit{post hoc} analysis. To our knowledge, ARGMINN is the most MI (yet still fully expressive) architecture for genome regulation---a field where \textit{understanding} a model's learned decision rules is equally as critical as its accuracy \citep{Eraslan2019,Rudin2019}---and is one of the few intrinsically MI architectures of its complexity in general. We showed that ARGMINN is expressive enough to accurately predict genome regulation, yet is uniquely constrained so the weights and activations directly encode the decision process \textit{in a human-interpretable way}. Further work in this area can yield major benefits for scientific AI and explainable AI.

\clearpage

\bibliography{mechintnet}

\begin{thebibliography}{46}
\providecommand{\natexlab}[1]{#1}
\providecommand{\url}[1]{\texttt{#1}}
\expandafter\ifx\csname urlstyle\endcsname\relax
  \providecommand{\doi}[1]{doi: #1}\else
  \providecommand{\doi}{doi: \begingroup \urlstyle{rm}\Url}\fi

\bibitem[Alipanahi et~al.(2015)Alipanahi, Delong, Weirauch, and
  Frey]{Alipanahi2015}
Babak Alipanahi, Andrew Delong, Matthew~T. Weirauch, and Brendan~J. Frey.
\newblock Predicting the sequence specificities of dna- and rna-binding
  proteins by deep learning.
\newblock \emph{Nature Biotechnology}, 33:\penalty0 831--838, 8 2015.
\newblock ISSN 15461696.
\newblock \doi{10.1038/nbt.3300}.

\bibitem[Alvarez-Melis \& Jaakkola(2018)Alvarez-Melis and
  Jaakkola]{Alvarez-Melis2018}
David Alvarez-Melis and Tommi~S. Jaakkola.
\newblock On the robustness of interpretability methods.
\newblock 6 2018.
\newblock URL \url{https://arxiv.org/abs/1806.08049v1}.

\bibitem[Arrieta et~al.(2020)Arrieta, Díaz-Rodríguez, Ser, Bennetot, Tabik,
  Barbado, Garcia, Gil-Lopez, Molina, Benjamins, Chatila, and
  Herrera]{BarredoArrieta2020}
Alejandro~Barredo Arrieta, Natalia Díaz-Rodríguez, Javier~Del Ser, Adrien
  Bennetot, Siham Tabik, Alberto Barbado, Salvador Garcia, Sergio Gil-Lopez,
  Daniel Molina, Richard Benjamins, Raja Chatila, and Francisco Herrera.
\newblock Explainable artificial intelligence (xai): Concepts, taxonomies,
  opportunities and challenges toward responsible ai.
\newblock \emph{Information Fusion}, 58:\penalty0 82--115, 6 2020.
\newblock ISSN 1566-2535.
\newblock \doi{10.1016/J.INFFUS.2019.12.012}.

\bibitem[Bailey et~al.(2015)Bailey, Johnson, Grant, and Noble]{Bailey2015}
Timothy~L Bailey, James Johnson, Charles~E Grant, and William~S Noble.
\newblock The meme suite.
\newblock \emph{Nucleic Acids Research}, 43:\penalty0 W39--49, 7 2015.
\newblock \doi{10.1093/nar/gkv416}.

\bibitem[Barbiero et~al.(2023)Barbiero, Ciravegna, Giannini, Zarlenga,
  Magister, Tonda, Lio, Precioso, Jamnik, and Marra]{Barbiero2023}
Pietro Barbiero, Gabriele Ciravegna, Francesco Giannini, Mateo~Espinosa
  Zarlenga, Lucie~Charlotte Magister, Alberto Tonda, Pietro Lio, Frederic
  Precioso, Mateja Jamnik, and Giuseppe Marra.
\newblock Interpretable neural-symbolic concept reasoning, 6 2023.

\bibitem[Benos et~al.(2002)Benos, Bulyk, and Stormo]{Benos2002}
Panayiotis~V. Benos, Martha~L. Bulyk, and Gary~D. Stormo.
\newblock Additivity in protein–dna interactions: how good an approximation
  is it?
\newblock \emph{Nucleic Acids Research}, 30:\penalty0 4442--4451, 10 2002.
\newblock ISSN 0305-1048.
\newblock \doi{10.1093/NAR/GKF578}.
\newblock URL \url{https://dx.doi.org/10.1093/nar/gkf578}.

\bibitem[Bereska \& Gavves(2024)Bereska and Gavves]{Bereska2024}
Leonard Bereska and Efstratios Gavves.
\newblock Mechanistic interpretability for ai safety -- a review.
\newblock 4 2024.
\newblock URL \url{https://arxiv.org/abs/2404.14082v2}.

\bibitem[Chiang et~al.(2023)Chiang, Cholak, and Pillay]{Chiang2023}
David Chiang, Peter Cholak, and Anand Pillay.
\newblock Tighter bounds on the expressivity of transformer encoders.
\newblock \emph{Proceedings of Machine Learning Research}, 202:\penalty0
  5544--5562, 1 2023.
\newblock ISSN 26403498.
\newblock URL \url{https://arxiv.org/abs/2301.10743v3}.

\bibitem[Consortium(2012)]{ENCODE_Project_Consortium2012-wb}
ENCODE~Project Consortium.
\newblock An integrated encyclopedia of dna elements in the human genome.
\newblock \emph{Nature}, 489:\penalty0 57--74, 9 2012.
\newblock \doi{10.1038/nature11247}.

\bibitem[Dosilovic et~al.(2018)Dosilovic, Brcic, and Hlupic]{Dosilovic2018}
Filip~Karlo Dosilovic, Mario Brcic, and Nikica Hlupic.
\newblock Explainable artificial intelligence: A survey.
\newblock \emph{2018 41st International Convention on Information and
  Communication Technology, Electronics and Microelectronics, MIPRO 2018 -
  Proceedings}, pp.\  210--215, 6 2018.
\newblock \doi{10.23919/MIPRO.2018.8400040}.

\bibitem[Eraslan et~al.(2019)Eraslan, Žiga Avsec, Gagneur, and
  Theis]{Eraslan2019}
Gökcen Eraslan, Žiga Avsec, Julien Gagneur, and Fabian~J. Theis.
\newblock Deep learning: new computational modelling techniques for genomics.
\newblock \emph{Nature Reviews Genetics 2019 20:7}, 20:\penalty0 389--403, 4
  2019.
\newblock ISSN 1471-0064.
\newblock \doi{10.1038/s41576-019-0122-6}.
\newblock URL \url{https://www.nature.com/articles/s41576-019-0122-6}.

\bibitem[Fornes et~al.(2020)Fornes, Castro-Mondragon, Khan, van~der Lee, Zhang,
  Richmond, Modi, Correard, Gheorghe, Baranašić, Santana-Garcia, Tan,
  Chèneby, Ballester, Parcy, Sandelin, Lenhard, Wasserman, and
  Mathelier]{Fornes2020}
Oriol Fornes, Jaime~A Castro-Mondragon, Aziz Khan, Robin van~der Lee, Xi~Zhang,
  Phillip~A Richmond, Bhavi~P Modi, Solenne Correard, Marius Gheorghe, Damir
  Baranašić, Walter Santana-Garcia, Ge~Tan, Jeanne Chèneby, Benoit
  Ballester, François Parcy, Albin Sandelin, Boris Lenhard, Wyeth~W Wasserman,
  and Anthony Mathelier.
\newblock Jaspar 2020: update of the open-access database of transcription
  factor binding profiles.
\newblock \emph{Nucleic Acids Research}, 48:\penalty0 D87--D92, 1 2020.
\newblock \doi{10.1093/nar/gkz1001}.

\bibitem[Friedman et~al.(2023)Friedman, Wettig, and Chen]{Friedman2023}
Dan Friedman, Alexander Wettig, and Danqi Chen.
\newblock Learning transformer programs.
\newblock 6 2023.
\newblock URL \url{https://arxiv.org/abs/2306.01128v2}.

\bibitem[Geusz et~al.(2021)Geusz, Wang, Lam, Vinckier, Alysandratos, Roberts,
  Wang, Kefalopoulou, Ramirez, Qiu, Chiou, Gaulton, Ren, Kotton, and
  Sander]{Geusz2021}
Ryan~J. Geusz, Allen Wang, Dieter~K. Lam, Nicholas~K. Vinckier,
  Konstantinos~Dionysios Alysandratos, David~A. Roberts, Jinzhao Wang, Samy
  Kefalopoulou, Araceli Ramirez, Yunjiang Qiu, Joshua Chiou, Kyle~J. Gaulton,
  Bing Ren, Darrell~N. Kotton, and Maike Sander.
\newblock Sequence logic at enhancers governs a dual mechanism of endodermal
  organ fate induction by foxa pioneer factors.
\newblock \emph{Nature Communications 2021 12:1}, 12:\penalty0 1--19, 11 2021.
\newblock ISSN 2041-1723.
\newblock \doi{10.1038/s41467-021-26950-0}.
\newblock URL \url{https://www.nature.com/articles/s41467-021-26950-0}.

\bibitem[Ghorbani et~al.(2017)Ghorbani, Abid, and Zou]{Ghorbani2017}
Amirata Ghorbani, Abubakar Abid, and James Zou.
\newblock Interpretation of neural networks is fragile.
\newblock \emph{33rd AAAI Conference on Artificial Intelligence, AAAI 2019,
  31st Innovative Applications of Artificial Intelligence Conference, IAAI 2019
  and the 9th AAAI Symposium on Educational Advances in Artificial
  Intelligence, EAAI 2019}, pp.\  3681--3688, 10 2017.
\newblock ISSN 2159-5399.
\newblock \doi{10.1609/aaai.v33i01.33013681}.
\newblock URL \url{https://arxiv.org/abs/1710.10547v2}.

\bibitem[Ghorbani et~al.(2019)Ghorbani, Brain, Zou, and Brain]{Ghorbani2019}
Amirata Ghorbani, James Wexler~Google Brain, James Zou, and Been Kim~Google
  Brain.
\newblock Towards automatic concept-based explanations.
\newblock \emph{Advances in Neural Information Processing Systems}, 32, 2019.
\newblock URL \url{https://github.com/amiratag/ACE}.

\bibitem[Gosai et~al.(2023)Gosai, Castro, Fuentes, Butts, Kales, Noche, Mouri,
  Sabeti, Reilly, and Tewhey]{Gosai2023}
SJ~Gosai, RI~Castro, N~Fuentes, JC~Butts, S~Kales, RR~Noche, K~Mouri,
  PC~Sabeti, SK~Reilly, and R~Tewhey.
\newblock Machine-guided design of synthetic cell type-specific cis-regulatory
  elements.
\newblock \emph{bioRxiv}, pp.\  2023.08.08.552077, 8 2023.
\newblock \doi{10.1101/2023.08.08.552077}.
\newblock URL \url{https://www.biorxiv.org/content/10.1101/2023.08.08.552077v1
  https://www.biorxiv.org/content/10.1101/2023.08.08.552077v1.abstract}.

\bibitem[Kasioumis et~al.(2021)Kasioumis, Townsend, and
  Inakoshi]{Kasioumis2021}
Theodoros Kasioumis, Joe Townsend, and Hiroya Inakoshi.
\newblock Elite backprop: Training sparse interpretable neurons.
\newblock \emph{International Workshop on Neural-Symbolic Learning and
  Reasoning}, 2021.

\bibitem[Kelley et~al.(2016)Kelley, Snoek, and Rinn]{Kelley2016}
David~R. Kelley, Jasper Snoek, and John~L. Rinn.
\newblock Basset: Learning the regulatory code of the accessible genome with
  deep convolutional neural networks.
\newblock \emph{Genome Research}, 26:\penalty0 990--999, 7 2016.
\newblock ISSN 15495469.
\newblock \doi{10.1101/gr.200535.115}.

\bibitem[Kim et~al.(2017)Kim, Wattenberg, Gilmer, Cai, Wexler, Viegas, and
  Sayres]{Kim2017}
Been Kim, Martin Wattenberg, Justin Gilmer, Carrie Cai, James Wexler, Fernanda
  Viegas, and Rory Sayres.
\newblock Interpretability beyond feature attribution: Quantitative testing
  with concept activation vectors (tcav).
\newblock \emph{35th International Conference on Machine Learning, ICML 2018},
  6:\penalty0 4186--4195, 11 2017.
\newblock URL \url{https://arxiv.org/abs/1711.11279v5}.

\bibitem[Kindermans et~al.(2017)Kindermans, Hooker, Adebayo, Alber, Schütt,
  Dähne, Erhan, and Kim]{Kindermans2017}
Pieter~Jan Kindermans, Sara Hooker, Julius Adebayo, Maximilian Alber,
  Kristof~T. Schütt, Sven Dähne, Dumitru Erhan, and Been Kim.
\newblock The (un)reliability of saliency methods.
\newblock \emph{Lecture Notes in Computer Science (including subseries Lecture
  Notes in Artificial Intelligence and Lecture Notes in Bioinformatics)}, 11700
  LNCS:\penalty0 267--280, 11 2017.
\newblock ISSN 16113349.
\newblock \doi{10.1007/978-3-030-28954-6_14}.
\newblock URL \url{https://arxiv.org/abs/1711.00867v1}.

\bibitem[Koh et~al.(2020)Koh, Nguye, Tang, Mussmann, Pierso, Kim, and
  Liang]{Koh2020}
Pang~Wei Koh, Thao Nguye, Yew~Siang Tang, Stephen Mussmann, Emma Pierso, Been
  Kim, and Percy Liang.
\newblock Concept bottleneck models.
\newblock \emph{37th International Conference on Machine Learning, ICML 2020},
  PartF168147-7:\penalty0 5294--5304, 7 2020.
\newblock URL \url{https://arxiv.org/abs/2007.04612v3}.

\bibitem[Lambert et~al.(2018)Lambert, Jolma, Campitelli, Das, Yin, Albu, Chen,
  Taipale, Hughes, and Weirauch]{Lambert2018-gq}
Samuel~A Lambert, Arttu Jolma, Laura~F Campitelli, Pratyush~K Das, Yimeng Yin,
  Mihai Albu, Xiaoting Chen, Jussi Taipale, Timothy~R Hughes, and Matthew~T
  Weirauch.
\newblock The human transcription factors.
\newblock \emph{Cell}, 172:\penalty0 650--665, 2 2018.
\newblock \doi{10.1016/j.cell.2018.01.029}.

\bibitem[Lee et~al.(2015)Lee, Gorkin, Baker, Strober, Asoni, McCallion, and
  Beer]{Lee2015}
Dongwon Lee, David~U. Gorkin, Maggie Baker, Benjamin~J. Strober, Alessandro~L.
  Asoni, Andrew~S. McCallion, and Michael~A. Beer.
\newblock A method to predict the impact of regulatory variants from dna
  sequence.
\newblock \emph{Nature Genetics 2015 47:8}, 47:\penalty0 955--961, 6 2015.
\newblock ISSN 1546-1718.
\newblock \doi{10.1038/ng.3331}.
\newblock URL \url{https://www.nature.com/articles/ng.3331}.

\bibitem[Lemon \& Tjian(2000)Lemon and Tjian]{Lemon2000}
Bryan Lemon and Robert Tjian.
\newblock Orchestrated response: a symphony of transcription factors for gene
  control.
\newblock \emph{Genes \& Development}, 14:\penalty0 2551--2569, 10 2000.
\newblock ISSN 0890-9369.
\newblock \doi{10.1101/GAD.831000}.
\newblock URL \url{http://genesdev.cshlp.org/content/14/20/2551.full
  http://genesdev.cshlp.org/content/14/20/2551}.

\bibitem[Linder et~al.(2023)Linder, Srivastava, Yuan, Agarwal, and
  Kelley]{Linder2023}
Johannes Linder, Divyanshi Srivastava, Han Yuan, Vikram Agarwal, and David~R.
  Kelley.
\newblock Predicting rna-seq coverage from dna sequence as a unifying model of
  gene regulation.
\newblock \emph{bioRxiv}, pp.\  2023.08.30.555582, 9 2023.
\newblock \doi{10.1101/2023.08.30.555582}.
\newblock URL \url{https://www.biorxiv.org/content/10.1101/2023.08.30.555582v1
  https://www.biorxiv.org/content/10.1101/2023.08.30.555582v1.abstract}.

\bibitem[Liu et~al.(2020)Liu, Xu, Rosenthal, juan Zhang, McCubbin, Meshgin,
  Shang, Koyama, Ma, Sharma, Heinz, Glass, Benner, Brenner, and
  Kisseleva]{Liu2020}
Xiao Liu, Jun Xu, Sara Rosenthal, Ling juan Zhang, Ryan McCubbin, Nairika
  Meshgin, Linshan Shang, Yukinori Koyama, Hsiao~Yen Ma, Sonia Sharma, Sven
  Heinz, Chris~K. Glass, Chris Benner, David~A. Brenner, and Tatiana Kisseleva.
\newblock Identification of lineage-specific transcription factors that prevent
  activation of hepatic stellate cells and promote fibrosis resolution.
\newblock \emph{Gastroenterology}, 158:\penalty0 1728--1744.e14, 5 2020.
\newblock ISSN 0016-5085.
\newblock \doi{10.1053/J.GASTRO.2020.01.027}.

\bibitem[Liu et~al.(2023)Liu, Gan, and Tegmark]{Liu2023}
Ziming Liu, Eric Gan, and Max Tegmark.
\newblock Seeing is believing: Brain-inspired modular training for mechanistic
  interpretability.
\newblock 5 2023.
\newblock URL \url{https://arxiv.org/abs/2305.08746v2}.

\bibitem[Novakovsky et~al.(2022{\natexlab{a}})Novakovsky, Dexter, Libbrecht,
  Wasserman, and Mostafavi]{Novakovsky2022r}
Gherman Novakovsky, Nick Dexter, Maxwell~W. Libbrecht, Wyeth~W. Wasserman, and
  Sara Mostafavi.
\newblock Obtaining genetics insights from deep learning via explainable
  artificial intelligence.
\newblock \emph{Nature Reviews Genetics 2022 24:2}, 24:\penalty0 125--137, 10
  2022{\natexlab{a}}.
\newblock ISSN 1471-0064.
\newblock \doi{10.1038/s41576-022-00532-2}.
\newblock URL \url{https://www.nature.com/articles/s41576-022-00532-2}.

\bibitem[Novakovsky et~al.(2022{\natexlab{b}})Novakovsky, Fornes, Saraswat,
  Mostafavi, and Wasserman]{Novakovsky2022}
Gherman Novakovsky, Oriol Fornes, Manu Saraswat, Sara Mostafavi, and Wyeth~W.
  Wasserman.
\newblock Explainn: interpretable and transparent neural networks for genomics.
\newblock \emph{bioRxiv}, pp.\  2022.05.20.492818, 11 2022{\natexlab{b}}.
\newblock \doi{10.1101/2022.05.20.492818}.
\newblock URL \url{https://www.biorxiv.org/content/10.1101/2022.05.20.492818v3
  https://www.biorxiv.org/content/10.1101/2022.05.20.492818v3.abstract}.

\bibitem[Pan \& Phan(2008)Pan and Phan]{Pan2008}
Youlian Pan and Sieu Phan.
\newblock Guide to threshold selection for motif prediction using positional
  weight matrix.
\newblock \emph{Proceedings of the International MultiConference of Engineers
  and Computer Scientists}, 1, 2008.

\bibitem[Riegel et~al.(2020)Riegel, Gray, Luus, Khan, Makondo, Akhalwaya, Qian,
  Fagin, Barahona, Sharma, Ikbal, Karanam, Neelam, Likhyani, and
  Srivastava]{Riegel2020}
Ryan Riegel, Alexander Gray, Francois Luus, Naweed Khan, Ndivhuwo Makondo,
  Ismail~Yunus Akhalwaya, Haifeng Qian, Ronald Fagin, Francisco Barahona, Udit
  Sharma, Shajith Ikbal, Hima Karanam, Sumit Neelam, Ankita Likhyani, and
  Santosh Srivastava.
\newblock Logical neural networks.
\newblock 6 2020.
\newblock URL \url{https://arxiv.org/abs/2006.13155v1}.

\bibitem[Rudin(2019)]{Rudin2019}
Cynthia Rudin.
\newblock Stop explaining black box machine learning models for high stakes
  decisions and use interpretable models instead.
\newblock \emph{Nature Machine Intelligence 2019 1:5}, 1:\penalty0 206--215, 5
  2019.
\newblock ISSN 2522-5839.
\newblock \doi{10.1038/s42256-019-0048-x}.
\newblock URL \url{https://www.nature.com/articles/s42256-019-0048-x}.

\bibitem[Schreiber et~al.(2020)Schreiber, Lu, and Noble]{Schreiber2020}
Jacob Schreiber, Yang~Young Lu, and William~Stafford Noble.
\newblock Ledidi: Designing genomic edits that induce functional activity.
\newblock \emph{bioRxiv}, pp.\  2020.05.21.109686, 5 2020.
\newblock \doi{10.1101/2020.05.21.109686}.
\newblock URL \url{https://www.biorxiv.org/content/10.1101/2020.05.21.109686v1
  https://www.biorxiv.org/content/10.1101/2020.05.21.109686v1.abstract}.

\bibitem[Seachrist et~al.(2021)Seachrist, Anstine, and Keri]{Seachrist2021}
Darcie~D. Seachrist, Lindsey~J. Anstine, and Ruth~A. Keri.
\newblock Foxa1: A pioneer of nuclear receptor action in breast cancer.
\newblock \emph{Cancers}, 13, 10 2021.
\newblock ISSN 20726694.
\newblock \doi{10.3390/CANCERS13205205}.
\newblock URL \url{/pmc/articles/PMC8533709/
  /pmc/articles/PMC8533709/?report=abstract
  https://www.ncbi.nlm.nih.gov/pmc/articles/PMC8533709/}.

\bibitem[Shrikumar et~al.(2017)Shrikumar, Greenside, and
  Kundaje]{Shrikumar2017}
Avanti Shrikumar, Peyton Greenside, and Anshul Kundaje.
\newblock Learning important features through propagating activation
  differences.
\newblock \emph{Proceedings of Machine Learning Research}, pp.\  3145--3153, 7
  2017.
\newblock ISSN 1938-7228.
\newblock URL \url{http://proceedings.mlr.press/v70/shrikumar17a.html}.

\bibitem[Shrikumar et~al.(2018)Shrikumar, Tian, Shcherbina, Žiga Avsec,
  Banerjee, Sharmin, Nair, and Kundaje]{Shrikumar2018}
Avanti Shrikumar, Katherine Tian, Anna Shcherbina, Žiga Avsec, Abhimanyu
  Banerjee, Mahfuza Sharmin, Surag Nair, and Anshul Kundaje.
\newblock Tf-modisco v0.4.2.2-alpha: Technical note.
\newblock \emph{arXiv}, 10 2018.
\newblock URL \url{http://arxiv.org/abs/1811.00416}.

\bibitem[Siggers \& Gordân(2014)Siggers and Gordân]{Siggers2014}
Trevor Siggers and Raluca Gordân.
\newblock Protein–dna binding: complexities and multi-protein codes.
\newblock \emph{Nucleic Acids Research}, 42:\penalty0 2099--2111, 2 2014.
\newblock ISSN 0305-1048.
\newblock \doi{10.1093/NAR/GKT1112}.

\bibitem[Stormo \& Fields(1998)Stormo and Fields]{Stormo1998}
Gary~D. Stormo and Dana~S. Fields.
\newblock Specificity, free energy and information content in protein–dna
  interactions.
\newblock \emph{Trends in Biochemical Sciences}, 23:\penalty0 109--113, 3 1998.
\newblock ISSN 0968-0004.
\newblock \doi{10.1016/S0968-0004(98)01187-6}.

\bibitem[Sundararajan et~al.(2017)Sundararajan, Taly, and
  Yan]{Sundararajan2017}
Mukund Sundararajan, Ankur Taly, and Qiqi Yan.
\newblock Axiomatic attribution for deep networks.
\newblock \emph{arXiv}, 3 2017.
\newblock URL \url{http://arxiv.org/abs/1703.01365}.

\bibitem[Tang et~al.(2021)Tang, Jia, Xu, Da, and Wu]{Tang2021}
Yuanxiao Tang, Zhilian Jia, Honglin Xu, Lin~Tai Da, and Qiang Wu.
\newblock Mechanism of rest/nrsf regulation of clustered protocadherin alpha
  genes.
\newblock \emph{Nucleic Acids Research}, 49:\penalty0 4506--4521, 5 2021.
\newblock ISSN 0305-1048.
\newblock \doi{10.1093/NAR/GKAB248}.
\newblock URL \url{https://dx.doi.org/10.1093/nar/gkab248}.

\bibitem[Tseng et~al.(2020)Tseng, Shrikumar, and Kundaje]{Tseng2020a}
Alex~M. Tseng, Avanti Shrikumar, and Anshul Kundaje.
\newblock Fourier-transform-based attribution priors improve the
  interpretability and stability of deep learning models for genomics.
\newblock \emph{bioRxiv}, pp.\  2020.06.11.147272, 6 2020.
\newblock \doi{10.1101/2020.06.11.147272}.
\newblock URL \url{https://www.biorxiv.org/content/10.1101/2020.06.11.147272v1
  https://www.biorxiv.org/content/10.1101/2020.06.11.147272v1.abstract}.

\bibitem[Tseng(2022)]{Tseng2022}
Alex~Michael Tseng.
\newblock Improving and leveraging the interpretability of deep neural networks
  for genomics, 2022.
\newblock URL \url{https://purl.stanford.edu/jv141vb2060}.

\bibitem[Vierstra et~al.(2020)Vierstra, Lazar, Sandstrom, Halow, Lee, Bates,
  Diegel, Dunn, Neri, Haugen, Rynes, Reynolds, Nelson, Johnson, Frerker,
  Buckley, Kaul, Meuleman, and Stamatoyannopoulos]{Vierstra2020}
Jeff Vierstra, John Lazar, Richard Sandstrom, Jessica Halow, Kristen Lee,
  Daniel Bates, Morgan Diegel, Douglas Dunn, Fidencio Neri, Eric Haugen, Eric
  Rynes, Alex Reynolds, Jemma Nelson, Audra Johnson, Mark Frerker, Michael
  Buckley, Rajinder Kaul, Wouter Meuleman, and John~A. Stamatoyannopoulos.
\newblock Global reference mapping of human transcription factor footprints.
\newblock \emph{Nature 2020 583:7818}, 583:\penalty0 729--736, 7 2020.
\newblock ISSN 1476-4687.
\newblock \doi{10.1038/s41586-020-2528-x}.
\newblock URL \url{https://www.nature.com/articles/s41586-020-2528-x}.

\bibitem[Žiga Avsec et~al.(2021{\natexlab{a}})Žiga Avsec, Agarwal, Visentin,
  Ledsam, Grabska-Barwinska, Taylor, Assael, Jumper, Kohli, and
  Kelley]{Avsec2021}
Žiga Avsec, Vikram Agarwal, Daniel Visentin, Joseph~R. Ledsam, Agnieszka
  Grabska-Barwinska, Kyle~R. Taylor, Yannis Assael, John Jumper, Pushmeet
  Kohli, and David~R. Kelley.
\newblock Effective gene expression prediction from sequence by integrating
  long-range interactions.
\newblock \emph{Nature Methods 2021 18:10}, 18:\penalty0 1196--1203, 10
  2021{\natexlab{a}}.
\newblock ISSN 1548-7105.
\newblock \doi{10.1038/s41592-021-01252-x}.
\newblock URL \url{https://www.nature.com/articles/s41592-021-01252-x}.

\bibitem[Žiga Avsec et~al.(2021{\natexlab{b}})Žiga Avsec, Weilert, Shrikumar,
  Krueger, Alexandari, Dalal, Fropf, McAnany, Gagneur, Kundaje, and
  Zeitlinger]{Avsec2021-yf}
Žiga Avsec, Melanie Weilert, Avanti Shrikumar, Sabrina Krueger, Amr
  Alexandari, Khyati Dalal, Robin Fropf, Charles McAnany, Julien Gagneur,
  Anshul Kundaje, and Julia Zeitlinger.
\newblock Base-resolution models of transcription-factor binding reveal soft
  motif syntax.
\newblock \emph{Nature Genetics}, 53:\penalty0 354--366, 3 2021{\natexlab{b}}.
\newblock \doi{10.1038/s41588-021-00782-6}.

\end{thebibliography}
\bibliographystyle{iclr2025_conference}

\clearpage

\appendix

\section{Supplementary Proofs}
\label{app:supp-proofs}

\subsection{Proof of Theorem \ref{thm:argminn-express}}

In this section, we prove that every configuration of motifs (which is definable by a first-order-logical sentence with position-indexed variables) is recognizable by an ARGMINN classifier. We will first show that ARGMINN’s motif-scanner module is sufficiently expressive to characterize motif-based protein binding, and then subsequently that the syntax-builder module can recognize arbitrary logical syntax between these binding sites.

\textbf{Biophysical assumptions}

We begin with our biophysical assumptions which justify our modeling of motif biology by first-order logic (with position-indexed variables).

\begin{enumerate}
    \item For binary biological readouts of interest (e.g., protein-binding measured by ChIP-seq, DNA accessibility measured by DNase-seq, etc.), the readout is fully characterized by the binding of proteins, which recognize motifs in the sequence \citep{Stormo1998,Lemon2000}.
    \item The free concentration of any particular protein is constant across training and testing conditions.
    \item For a given potential binding site (e.g., a motif instance in a DNA sequence), the strength and likelihood of binding of a specific protein (these quantities are related through statistical mechanics) can be sufficiently summarized by a single scalar value (related to the $K_{d}$, or dissociation constant). Through statistical mechanics, the fraction/probability of binding is $\frac{[\text{TF}]}{K_{d} + [\text{TF}]}$ \citep{Stormo1998}.
    \item The $K_{d}$ value (or a monotonic function of it) can be computed as an independently additive function of individual positions of the binding site. An example of this is the PWM (position weight matrix), which has been shown to be a good approximation of binding mechanics \citep{Benos2002,Pan2008}.
    \item Input sequences are a finite length $\ell$.
    \item Given real-valued variables $b_{m,p}$ representing the binding strength of motif $m$ at position $p$ in a sequence, the binary biological readout (e.g., binding of a particular protein of interest or accessibility) of the sequence as a whole can be expressed as the following disjunction of statements:
    
    $\sigma := \bigvee\limits_{i=1}^{n}\phi_{i}$,
    
    where $\phi_{i}$ is a statement denoting a \textit{single} possible configuration of motifs that induces a positive biological readout.
    
    Each $\phi_{i}$ has the following form:
    
    $\phi_{i} := (b_{m_{i,1},p_{i,1}} \geq t_{i,1}) \land \cdots \land (b_{m_{i,d},p_{i,d}} \geq t_{i,d})$.
    
    That is, $\phi_{i}$ defines whether or not a specific configuration of $d \geq 1$ binding motifs $m_{i,1},…,m_{i,d}$ exist with sufficient strength at positions $p_{i,1},…,p_{i,d}$. Note that with finite sequences, all possible statements about combinations of motifs (expressable in first-order logic) can be written in this form, by the Disjunctive Normal Form Theorem.
\end{enumerate}

Below, we proceed with our proof by constructing an instantiation of the ARGMINN architecture where the particular instantiation implements/recognizes a sentence $\sigma$.

\textbf{Motif-scanner module}

Given the above biophysical assumptions, the motif-scanner module’s convolutional filters are sufficiently expressive to capture the binding strengths/likelihoods of each position for each potential binding motif.

We simply design this module to have $n_{f}$ filters equal to the number of unique motifs $m$ in $\sigma$, with width $w$ equal to the maximum width of any motif. We set the multiplicative weights of each filter with the PWM of each motif, so that this module precisely implements the PWM-scanning procedure that is common in the field of regulatory genomics \citep{Benos2002}. For this proof, we set the bias to be 0. Note that if the convolutional filters $W$ are all PWMs, then the multiplicative weights are all non-negative, and so the pre-ReLU activation will also be non-negative.

Thus, the motif-scanner module outputs a set of binding strengths for each position of the sequence, for each possible motif. In other words, this module outputs $b_{m,p}$ for each relevant motif $m$, at each position $p$. This constitutes the motif activations $A$ of the model.

\textbf{Syntax-builder module}

Now we design an instantiation of the syntax-builder module which captures the logic in $\sigma$. In particular, our model will produce a pre-sigmoid output which is positive if and only if $\sigma$ is true.

The input to the syntax-builder module is the concatenation of motif activations $A$ (which is the matrix of binding strengths $b_{m,p}$) and the positional encodings $P$: $A\Vert P$. For ease of proof, let the positional encodings $P$ be a one-hot-encoded vector of size $\ell$, denoting position. We will use the notation $[A\Vert P]_{p}$ to denote the column vector at position $p$, which is an $(n_{f} + \ell)$-dimensional vector where the first $n_{f}$ entries contain the binding strengths/activations of each motif at position $p$, and the latter $\ell$ entries contain a one-hot encoding of position $p$.

Recall that within $\sigma$, $n$ is the number of statements in disjunction. Let $d_{max}$ be the maximum length of any of the $\phi_{i}$s. For simplicity of our proof and construction, we pad every $\phi_{i}$ to have exactly $d_{max}$ clauses by adding ``dummy motif clauses''. For example, let us pad with the clause $(b_{0,0} \geq 0)$. This simply checks that some arbitrary motif (index 0) at some arbitrary position (position 0) has non-negative activation, which will always be true since the convolutional weights in the motif-scanner module are PWMs.

We let the memory vector $m_{l}$ be of dimension $n(n_{f} + \ell)$. Additionally, we define the dimension of the query/key/value vectors be $n_{f} + \ell$.

Our proof (like many other similar proofs) relies on the universal approximation theorem of feed-forward networks (FFNs). We also leverage Lemma 19 from \citep{Chiang2023}, which shows that in an FFN, residual connections can effectively be ignored (which simplifies the construction of our network).

We proceed with induction on $d_{max}$, where our syntax-builder module has $d_{max}$ attention layers, with $n$ heads for each layer (one head for every $\phi_{i}$).

\underline{Base case: $d_{max} = 1$}

Consider $d_{max} = 1$. In this case, each $\phi_{i}$ can be written as $\phi_{i} := b_{m_{i},p_{i}} \geq t_{i}$. That is, $\phi_{i}$ is true if and only if motif $m_{i}$ at position $p_{i}$ is strong enough.

Now we instantiate a single-attention-layer syntax-builder module with $n$ attention heads. We repeat the following for each head independently:

For each head $i$, we learn to recognize $\phi_{i}$. Recall that the memory vector $m_{0}$ is initialized to be all 1s.

We define the weight matrices $W_{Q,1},W_{K,1},W_{V,1}$ separately for each attention head in this proof, knowing that the final weight matrices for the attention layer as a whole are obtained by a simple concatenatation operation over the heads. Let $W_{Q,1}$ map $m_{0}$ to a single vector where the first $n_{f}$ entries is a one-hot encoding where the 1 is in the position of motif $m_{i}$, and the latter $\ell$ entries is filled with a very negative constant $-C$ ($C >> 0$), except for the position $p_{i}$ (indexed from within these latter $\ell$ entries), which has a 1. Let $W_{K,1}=W_{V,1} = I_{n_{f} + \ell}$, the identity matrix of appropriate size:

$$q_{1} = W_{Q,1}m_{0} = \left[\begin{array}{*{15}{c}}0 & \cdots & 0 & 1 & 0 & \cdots & 0 & \vert & -C & \cdots & -C & 1 & -C & \cdots & -C \end{array}\right]^{\intercal}$$
where the 1 is at index $m_{i}$ in the left block, and the 1 is at index $p_{i}$ within the right block.

$$K_{1} = W_{K,1}[A\Vert P] = [A\Vert P] \quad V_{1} = W_{V,1}[A\Vert P] = [A\Vert P]$$

Next, we take the dot product of every key vector with the query vector. For the key vector $[A\Vert P]_{p_{i}}$ (originating from position $p_{i}$), the dot product will be $b_{m_{i},p_{i}} + 1$ (the first $n_{f}$ entries contribute the binding strength $b_{m_{i},p_{i}}$, and the latter $\ell$ entries contribute the 1). For any other position $p \neq p_{i}$, the dot product will be $b_{m_{i},p} - C$.

We select $C$ to be a large-enough magnitude such that after the softmax, the attention score at position $p_{i}$ will be approximately 1, and all other positions will effectively be 0. Thus, the final vector being passed to the feed-forward network (FFN) is equivalent to the value vector corresponding to position $p_{i}$ (which is equivalent to the motif activations/positional encoding at position $p_{i}$):

$$a_{1}V_{1} = [A\Vert P]_{p_{i}}$$

With all attention heads together, we obtain a concatenated vector of size $n(n_{f} + \ell)$, where every contiguous $i$th block of $n_{f} + \ell$ entries corresponds to $\phi_{i}$, and contains the vector $[A\Vert P]_{p_{i}}$.

We then design our FFN to map from this vector to our final memory stream $m_{1}$. In particular, this FFN will produce an output vector of the same size, where every contiguous $i$th block of $n_{f} + \ell$ entries contains all $-\frac{1}{n_{f} + \ell}$ if $b_{m_{i},p_{i}} < t_{i}$, and all $\frac{n}{n_{f} + \ell}$ otherwise. We invoke Lemma 19 from \citet{Chiang2023} and the Universal Approximation Theorem to perform this step.

Our final linear projection layer (which takes $m_{1}$ and maps to an output prediction $\hat{y}$) has weights of all 1 and bias of 0.

Together, this ensures that the output $\hat{y} > 0$ if and only if there exists an $i$ such that $b_{m_{i},p_{i}} > t_{i}$ (i.e., $\phi_{i}$ is true).

\underline{Base case: $d_{max} = 2$}

We proceed with a similar structure as with the above base case. In the first layer, we define $W_{Q,1},W_{K,1},W_{V,1}$ identically as above. However, after the first attention layer, we design the FFN differently so that $m_{1}$ contains information about the \textit{next} motif within $\phi_{i}$ to search for.

In particular, each $\phi_{i}$ is of the form $\phi_{i} := (b_{m_{i,1},p_{i,1}} \geq t_{i,1}) \land (b_{m_{i,2}, p_{i,2}} \geq t_{i,2})$. In the first attention layer, we define the weight matrices $W_{Q,1},W_{K,1},W_{V,1}$ as above, so that the vector passed to the FFN is a concatenation of $[A\Vert P]_{p_{i,1}}$ for all $i$.

Here, we design the FFN so that it will produce $m_{1}$, where every contiguous $i$th block of $n_{f} + \ell$ entries contains all 1s if $b_{m_{i,1},p_{i,1}} \geq t_{i,1}$, and all 0s otherwise.

Next, the second attention layer will identify the second motif in each $\phi_{i}$. Again, we consider each attention head $i$ separately.

In this second layer, $W_{Q,2}$ produces a query vector for each head which is similar to that in the previous base case: a vector of $n_{f} + \ell$ entries where the first $n_{f}$ is all 0 except for the position of $m_{i,2}$, and the latter $\ell$ entries are all $-C$ except for the position $p_{i,2}$ (however, note that if the first motif $m_{i,1}$ was not found at position $p_{i,1}$, then the $i$th block of $m_{1}$ will be all 0s, and so the query vector will be also all 0s). Again, we let $W_{K,2}=W_{V,2}=I_{n_{f} + \ell}$:

$$q_{2} = W_{Q,2}m_{1} = \left[\begin{array}{*{15}{c}}0 & \cdots & 0 & 1 & 0 & \cdots & 0 & \vert & -C & \cdots & -C & 1 & -C & \cdots & -C \end{array}\right]^{\intercal}$$
where the 1 is at index $m_{i,2}$ in the left block, and the 1 is at index $p_{i,2}$ within the right block. If the first motif $m_{i,1}$ was not found, then this vector will be all 0s.

$$K_{2} = W_{K,2}[A\Vert P] = [A\Vert P] \quad V_{2} = W_{V,2}[A\Vert P] = [A\Vert P]$$

We then follow the same construction as with the previous base case, where the second FFN produces the final memory stream $m_{2}$ based on comparing each $b_{m_{i,2}, p_{i,2}}$ to $t_{i,2}$: every contiguous $i$th block of $n_{f} + \ell$ entries contains all $-\frac{1}{n_{f} + \ell}$ if $b_{m_{i,2},p_{i,2}} < t_{i,2}$, and all $\frac{n}{n_{f} + \ell}$ otherwise. If, however, the first motif $m_{i,1}$ was not found, then the FFN will always output $-\frac{1}{n_{f} + \ell}$ for that block (in this case, the query vector is all 0s, and the latter $\ell$ entries of the $i$th block in the input to the FFN will be a smeared fraction rather than a one-hot encoding). The final projection layer will be the same as with the first base case, leading to the desired outcome.

\underline{Inductive case}
\nopagebreak

We complete our inductive proof for a general $d_{max}$, as the construction of the architecture with $d_{max}$ layers is a straightforward extension from the base cases.

We assume that in the first $d_{max} - 1$ layers, the memory stream $m_{d_{max} - 1}$ is structured as contiguous blocks of $n_{f} + \ell$ entries, where the $i$th block is such that it contains all 1s if and only if $b_{m_{i,j},p_{i,j}} \geq t_{i,j}$ for all $j < d_{max}$ and all 0s otherwise. We then structure our final $d_{max}$th layer of the attention mechanism similarly to the second layer in the base case of $d_{max} = 2$.

\subsection{Proof of Corollary \ref{corr:explainn-express}}

Here, we prove that there exists a configuration of motifs specified by first-order logic, which is not recognizable by ExplaiNN.

Consider a dataset of sequences defined by the presence of distinct motifs $A,B,C,D$. Every sequence has exactly two instances of such motifs. A positive sequence is defined by having both $A$ and $B$, or both $C$ and $D$. A negative sequence is defined by any other combination: $AC$, $BD$, $AD$, or $BC$. This constructed example is a realistic one, as it is the binding rule seen in transcription factors with two half sites (e.g., REST \citep{Tang2021}), or co-factor binding where the motifs and transcription factors are unidirectional (e.g., JUND and TEAD).

Within ExplaiNN, each CNN ``unit'' learns the presence of one such motif, and returns a scalar score. Let there be four CNN units, one for each motif. For each input sequence, we obtain four such scores: $s_{A},s_{B},s_{C},s_{D}$.

Each CNN unit consists of a single convolutional filter whose activation is maximized by the motif it learns. For simplicity, we assume that given a convolutional filter that learns substring $X$, the distribution of activations of that filter on background sequences is identical to the distribution of activations on non-$X$ motifs. This could be realized, for example, by motifs whose composition is base/letter-wise distributed identically to the background (e.g., uniform). Thus, for a convolutional filter which recognizes $X$, the distribution of its activations is identical across all non-$X$ substrings. Given sufficiently long sequences, the CNN unit for some motif $X$ will output a scalar score as follows: if $X$ is in the sequence, $s_{X} = p_{X}$, a score for ``positives''; if $X$ is not present, $s_{X} = n_{X} \neq p_{X}$, a score for ``negatives''.

Given these CNN units, suppose it is possible to distinguish positive and negative sequences with a linear combination, as in ExplaiNN. The output of the model is $w_{A}s_{A} + w_{B}s_{B} + w_{C}s_{C} + w_{D}s_{D} + \beta$ for scalar weights $w_{A},\ldots,w_{D}$ and bias $\beta$.

For the classifier to be sufficiently expressive, we require that positive examples have a final output that is at least some threshold $\tau$, and negative examples to have an output that is strictly less than $\tau$.

Thus, we have the following inequalities (one for each possible pair of motifs):

$$w_{A}p_{A} + w_{B}p_{B} + w_{C}n_{C} + w_{D}n_{D} + \beta \geq \tau$$
$$w_{A}n_{A} + w_{B}n_{B} + w_{C}p_{C} + w_{D}p_{D} + \beta \geq \tau$$
$$w_{A}p_{A} + w_{B}n_{B} + w_{C}p_{C} + w_{D}n_{D} + \beta < \tau$$
$$w_{A}p_{A} + w_{B}n_{B} + w_{C}n_{C} + w_{D}p_{D} + \beta < \tau$$
$$w_{A}n_{A} + w_{B}p_{B} + w_{C}p_{C} + w_{D}n_{D} + \beta < \tau$$
$$w_{A}n_{A} + w_{B}p_{B} + w_{C}n_{C} + w_{D}p_{D} + \beta < \tau$$

Adding inequalities of the same type:

$$w_{A}(p_{A} + n_{A}) + \cdots + w_{D}(p_{D} + n_{D}) \geq 2(\tau - \beta)$$
$$2w_{A}(p_{A} + n_{A}) + \cdots + 2w_{D}(p_{D} + n_{D}) < 4(\tau - \beta)$$

This is a contradiction, as this requires $w_{A}(p_{A} + n_{A}) + \cdots + w_{D}(p_{D} + n_{D})$ to be both at least $2(\tau-\beta)$ and strictly less than $2(\tau-\beta)$.

Thus, ExplaiNN is not sufficiently expressive to capture every configuration of motifs expressable in first-order logic (e.g., exclusive disjunctions).

\newpage
\clearpage

\section{Supplementary Figures and Tables}

\setcounter{table}{0}
\renewcommand{\thetable}{S\arabic{table}}
\setcounter{figure}{0}
\renewcommand{\thefigure}{S\arabic{figure}}

\begin{figure}[h]
\centering
\includegraphics[width=\columnwidth]{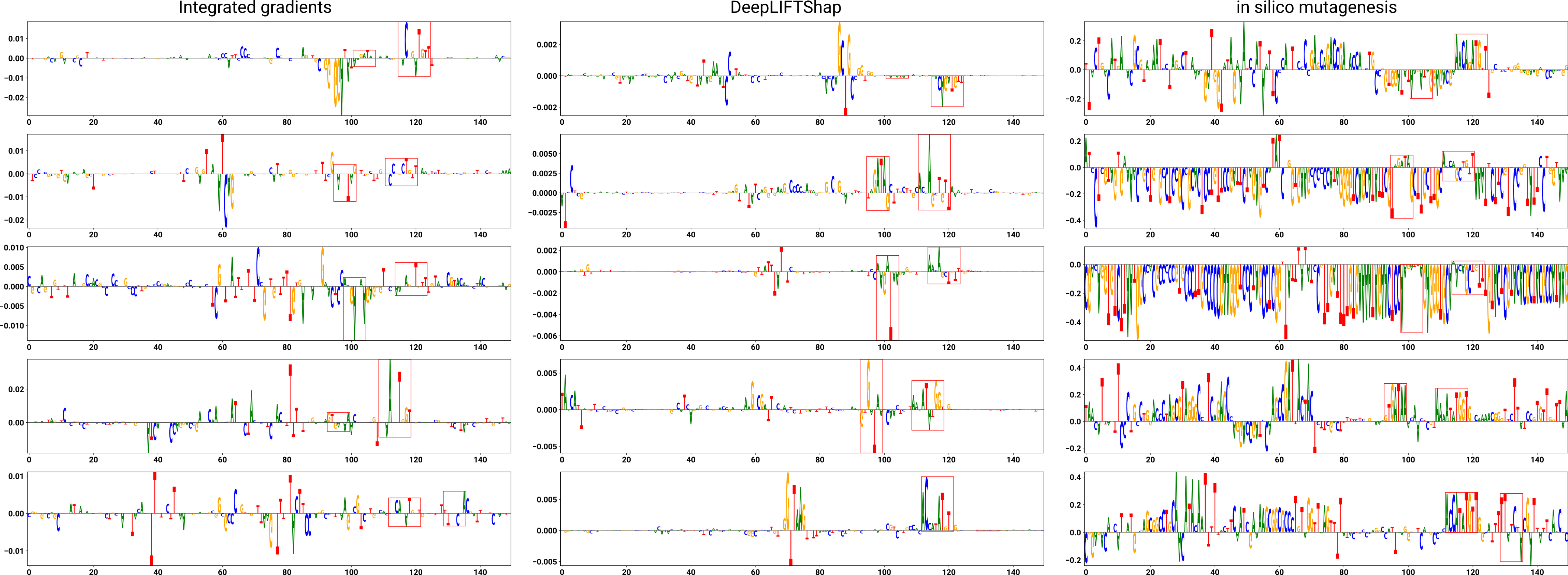}
\caption{\small From a single standard CNN model trained on TAL/GATA binding, we show the attribution scores for the same five input sequences, computed using three different methods: integrated gradients (left), DeepLIFTShap (middle), and \textit{in silico} mutagenesis (right). The locations of the true motifs are highlighted by the red box in each example. Although this model achieves near-perfect test accuracy, the importance scores remain unreliable and noisy. Regardless of the method, it is difficult to even identify \textit{where} the motif is solely based on these score tracks, let alone \textit{what} the motif is. Additionally, the methods disagree heavily with each other, even showing different signs (positive vs. negative) in importance for the true motifs.}
\label{suppfig:impscores}
\end{figure}

\begin{figure}[h]
\centering
\includegraphics[width=0.7\columnwidth]{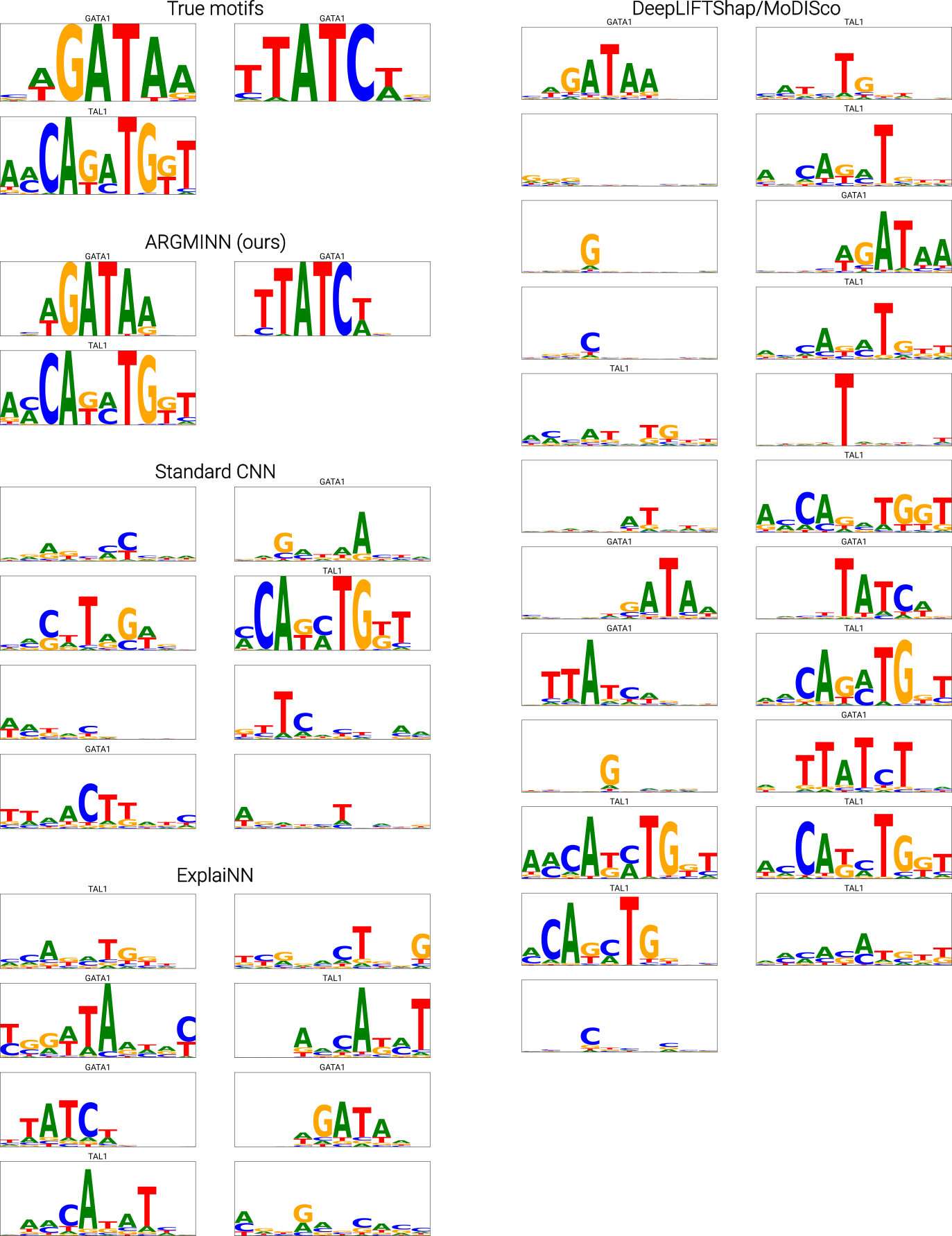}
\caption{\small On a simulated TAL/GATA dataset, we show all motifs discovered by: 1) ARGMINN, 2) interpreting the first-layer filters of a standard CNN, 3) ExplaiNN, and 4) running MoDISco on DeepLIFTShap importance scores. Each motif is labeled with the most similar motif from the simulation, using TOMTOM. Motifs which are not sufficiently similar to any of the motifs in the simulation (as determined by TOMTOM’s default thresholds), remain unlabeled. We also show the true motifs used in the simulation.}
\label{suppfig:more-disc-motifs-1}
\end{figure}

\begin{figure}[h]
\centering
\includegraphics[width=0.7\columnwidth]{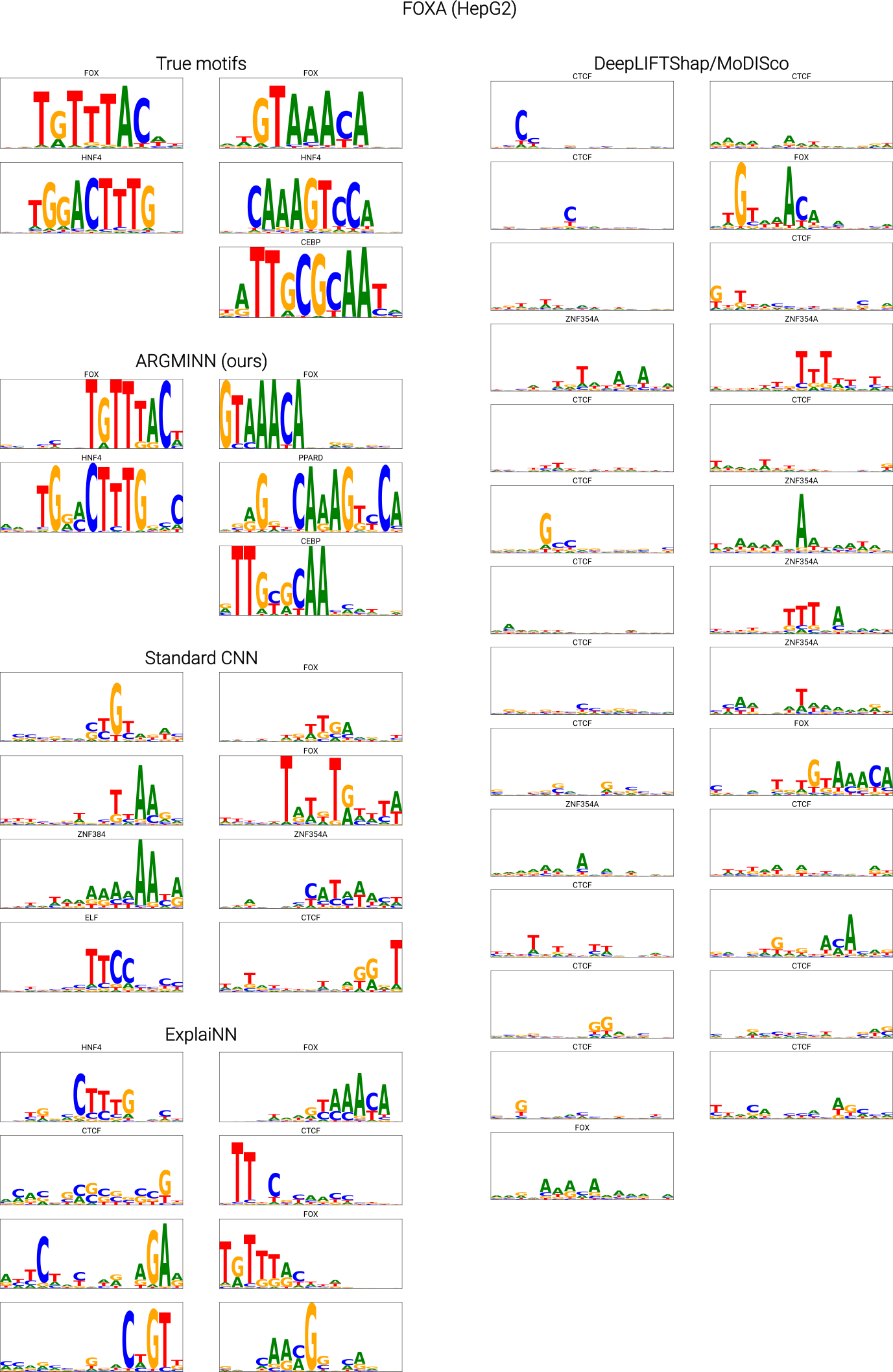}
\caption{\small On an experimental dataset of FOXA2 binding in HepG2, we show all motifs discovered by: 1) ARGMINN, 2) interpreting the first-layer filters of a standard CNN, 3) ExplaiNN, and 4) running MoDISco on DeepLIFTShap importance scores. Each motif is labeled with the most similar known human motif, using TOMTOM. Motifs which are not sufficiently similar to any known human motif (as determined by TOMTOM’s default thresholds), remain unlabeled. We also show ground-truth motifs from JASPAR which are supported by external literature.}
\label{suppfig:more-disc-motifs-2}
\end{figure}

\begin{figure}[h]
\centering
\includegraphics[width=0.9\columnwidth]{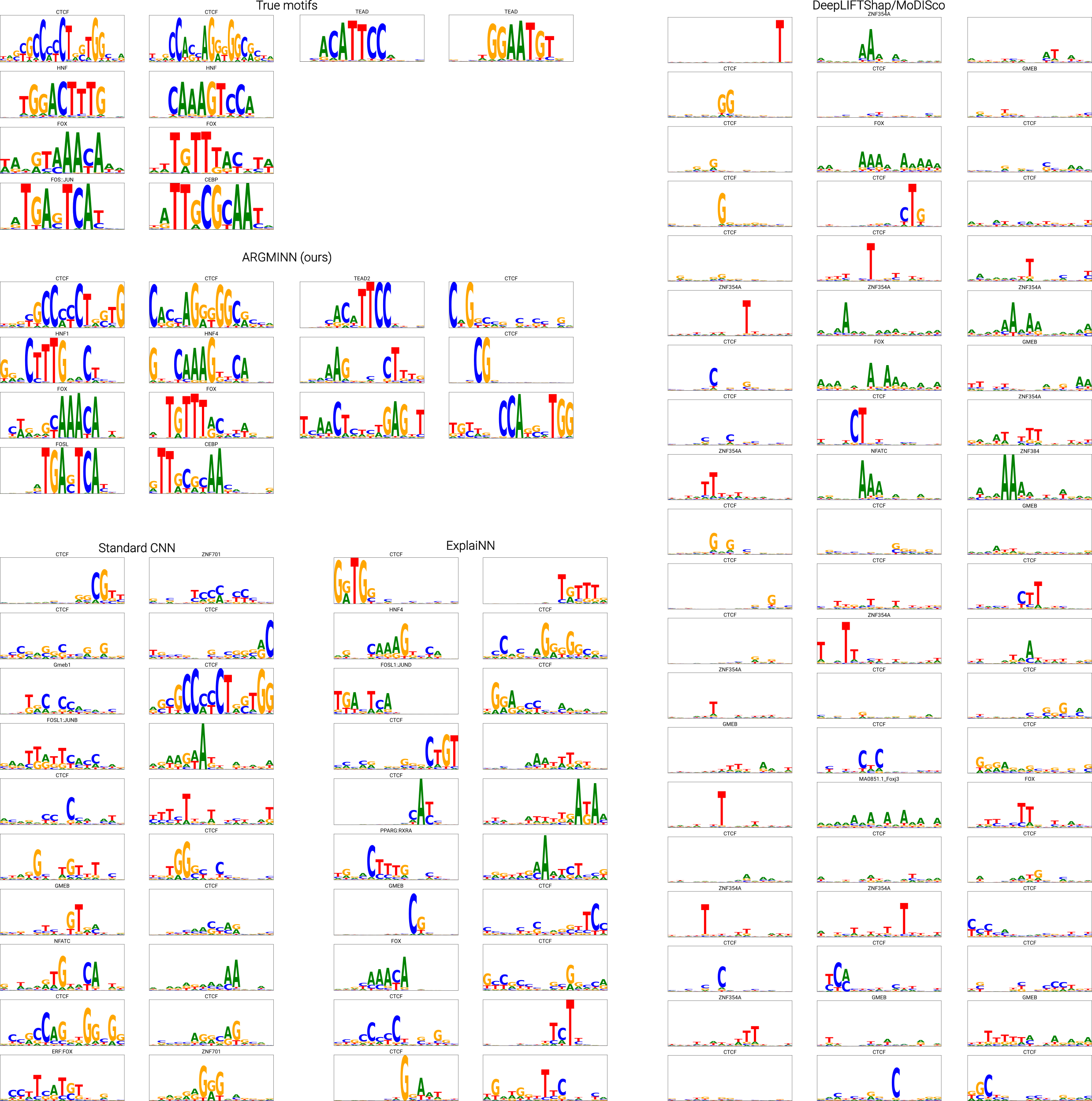}
\caption{\small On an experimental dataset of DNase accessibility in HepG2, we show all motifs discovered by: 1) ARGMINN, 2) interpreting the first-layer filters of a standard CNN, 3) ExplaiNN, and 4) running MoDISco on DeepLIFTShap importance scores. Each motif is labeled with the most similar known human motif, using TOMTOM. Motifs which are not sufficiently similar to any known human motif (as determined by TOMTOM’s default thresholds), remain unlabeled. We also show ground-truth motifs from JASPAR which are supported by external literature.}
\label{suppfig:more-disc-motifs-3}
\end{figure}

\begin{figure}[h]
\centering
\includegraphics[width=0.9\columnwidth]{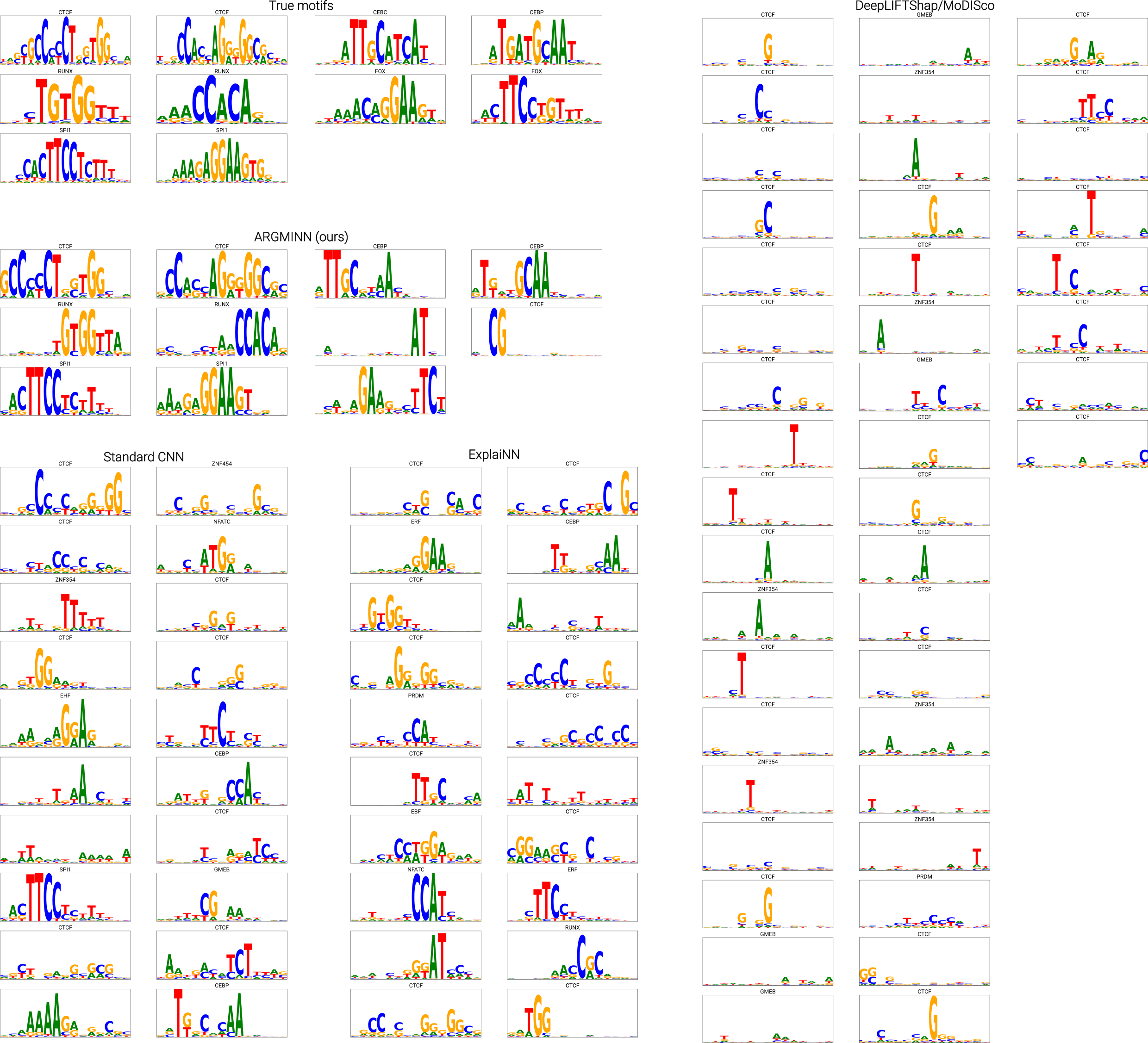}
\caption{\small On an experimental dataset of DNase accessibility in HL-60, we show all motifs discovered by: 1) ARGMINN, 2) interpreting the first-layer filters of a standard CNN, 3) ExplaiNN, and 4) running MoDISco on DeepLIFTShap importance scores. Each motif is labeled with the most similar known human motif, using TOMTOM. Motifs which are not sufficiently similar to any known human motif (as determined by TOMTOM’s default thresholds), remain unlabeled. We also show ground-truth motifs from JASPAR which are supported by external literature.}
\label{suppfig:more-disc-motifs-4}
\end{figure}

\FloatBarrier

\begin{table}[h]
    \caption{Identification and redundancy of discovered motifs}
    \label{supptab:motif-counts}
    \centering
    \small
    \begin{tabular}{llcccc}
        \toprule
        Dataset & Relevant motif & ARGMINN & Traditional CNN & ExplaiNN & DeepLIFTShap/MoDISco \\
        \midrule
        \multirow{1}{*}{SPI1} & SPI1 & \textbf{1} & 4 & 2 & 17 \\
        \multirow{2}{*}{TAL/GATA} & TAL & \textbf{1} & \textbf{1} & 3 & 10 \\
         & GATA1 & \textbf{1} & \textbf{1} & 2 & 3 \\
        \multirow{2}{*}{E2F6} & E2F6 & \textbf{1} & 2 & 3 & 10 \\
         & MAX & \textbf{1} & \textbf{1} & 2 & 5 \\
        \multirow{2}{*}{JUND/TEAD} & JUND-TRE & \textbf{1} & 3 & 3 & 22 \\
         & JUND-CRE & \textbf{1} & \textbf{1} & 2 & 5 \\
        \multirow{2}{*}{REST} & REST-left & \textbf{1} & \textbf{1} & \textbf{1} & 9 \\
         & REST-right & \textbf{1} & \textbf{1} & 3 & 4 \\
        \multirow{2}{*}{SPI1/CTCF} & SPI1 & \textbf{1} & 2 & 2 & 11 \\
         & CTCF & \textbf{1} & \textbf{1} & 2 & 2 \\
        \multirow{1}{*}{CTCF (HepG2)} & CTCF & 2 & 4 & 3 & 12 \\
        \multirow{3}{*}{FOXA1 (HepG2)} & FOX & \textbf{1} & 3 & \textbf{1} & 7 \\
         & HNF4 & \textbf{1} & 0 & \textbf{1} & 0 \\
         & CEBP & \textbf{1} & 0 & \textbf{1} & 0 \\
        \multirow{6}{*}{DNase (HepG2)} & HNF & 2 & 0 & 2 & 0 \\
         & TEAD & \textbf{1} & 0 & 0 & 0 \\
         & CTCF & 3 & 7 & 6 & 25 \\
         & FOX & \textbf{1} & 3 & \textbf{1} & 7 \\
         & CEBP & \textbf{1} & 0 & 0 & 0 \\
         & FOS::JUN & \textbf{1} & 0 & \textbf{1} & 0 \\
        \multirow{5}{*}{DNase (HL-60)} & RUNX & \textbf{1} & 0 & \textbf{1} & 0 \\
         & CTCF & 2 & 6 & 9 & 17 \\
         & FOX & 0 & 2 & \textbf{1} & 0 \\
         & CEBP & 2 & \textbf{1} & 2 & 0 \\
         & SPI1 & \textbf{1} & \textbf{1} & 0 & 0 \\
        \multirow{3}{*}{DNase (K562)} & FOSL2::JUN & \textbf{1} & 2 & \textbf{1} & 0 \\
         & CTCF & 3 & 8 & 7 & 8 \\
         & GATA & \textbf{1} & \textbf{1} & 2 & \textbf{1} \\
        \bottomrule
    \end{tabular}\\
    Number of times each relevant motif was discovered by each method. A value of 0 means the motif was not discovered at all. Values greater than 1 indicate redundancy.
\end{table}

\begin{table}[h]
    \caption{Similarity of discovered motifs to ground truth}
    \label{supptab:motif-sims}
    \centering
    \small
    \begin{tabular}{llcccc}
        \toprule
        Dataset & Relevant motif & ARGMINN & Traditional CNN & ExplaiNN & DeepLIFTShap/MoDISco \\
        \midrule
        \multirow{1}{*}{SPI1} & SPI1 & \textbf{10.890} & 1.583 & 4.876 & 8.452 \\
        \multirow{2}{*}{TAL/GATA} & TAL & 8.263 & 7.017 & 2.769 & \textbf{8.395} \\
         & GATA1 & \textbf{7.914} & 1.813 & 4.140 & 5.737 \\
        \multirow{2}{*}{E2F6} & E2F6 & \textbf{8.031} & 6.547 & 5.335 & 6.699 \\
         & MAX & \textbf{4.289} & 0.713 & 1.969 & 3.706 \\
        \multirow{2}{*}{JUND/TEAD} & JUND-TRE & 7.164 & 2.616 & 4.584 & \textbf{7.298} \\
         & JUND-CRE & 6.150 & 1.544 & 2.667 & \textbf{7.074} \\
        \multirow{2}{*}{REST} & REST-left & \textbf{13.727} & 0.428 & 1.783 & 8.027 \\
         & REST-right & \textbf{12.895} & 0.742 & 3.991 & 10.739 \\
        \multirow{2}{*}{SPI1/CTCF} & SPI1 & \textbf{10.339} & 6.894 & 6.938 & 8.578 \\
         & CTCF & \textbf{13.129} & 2.140 & 7.360 & 4.210 \\
        \multirow{1}{*}{CTCF (HepG2)} & CTCF & \textbf{21.974} & 6.800 & 15.945 & 13.338 \\
        \multirow{3}{*}{FOXA1 (HepG2)} & FOX & \textbf{5.651} & 2.712 & 4.917 & 5.017 \\
         & HNF4 & \textbf{7.188} & 0 & 4.587 & 0 \\
         & CEBP & \textbf{5.688} & 0 & 1.083 & 0 \\
        \multirow{6}{*}{DNase (HepG2)} & HNF & \textbf{6.178} & 0 & 5.852 & 0 \\
         & TEAD & \textbf{8.545} & 0 & 0 & 0 \\
         & CTCF & \textbf{25.482} & 24.543 & 19.120 & 4.262 \\
         & FOX & \textbf{6.831} & 1.270 & 3.815 & 2.028 \\
         & CEBP & \textbf{4.797} & 0 & 0 & 0 \\
         & FOS::JUN & \textbf{9.686} & 0 & 1.982 & 0 \\
        \multirow{5}{*}{DNase (HL-60)} & RUNX & \textbf{4.660} & 0 & 3.232 & 0 \\
         & CTCF & \textbf{24.428} & 6.377 & 16.945 & 5.423 \\
         & FOX & 0 & 1.149 & \textbf{4.178} & 0 \\
         & CEBP & \textbf{4.947} & 4.736 & 4.168 & 0 \\
         & SPI1 & \textbf{11.434} & 7.568 & 0 & 0 \\
        \multirow{3}{*}{DNase (K562)} & FOSL2::JUN & \textbf{7.646} & 7.338 & 5.198 & 0 \\
         & CTCF & \textbf{23.810} & 6.121 & 13.568 & 1.776 \\
         & GATA & \textbf{9.159} & 0.533 & 7.078 & 0.345 \\
        \bottomrule
    \end{tabular}\\
    Similarity of closest motif discovered by each method to each relevant motif. If a method did not discover a motif, it is given a similarity of 0.
\end{table}

\begin{table}[h]
    \caption{Extraneous discovered motifs}
    \label{supptab:extra-motifs}
    \centering
    \small
    \begin{tabular}{lcccc}
        \toprule
        Dataset & ARGMINN & Traditional CNN & ExplaiNN & DeepLIFTShap/MoDISco \\
        \midrule
        SPI1 & \textbf{0} & 1 & 4 & 10 \\
        TAL/GATA & \textbf{0} & 5 & 2 & 7 \\
        E2F6 & \textbf{0} & 3 & 2 & 4 \\
        JUND/TEAD & \textbf{0} & 4 & 3 & 2 \\
        REST & \textbf{0} & 6 & 1 & 7 \\
        SPI1/CTCF & \textbf{0} & 3 & 1 & 13 \\
        CTCF (HepG2) & \textbf{0} & 1 & 3 & 17 \\
        FOXA1 (HepG2) & \textbf{2} & 3 & 3 & 18 \\
        DNase (HepG2) & 3 & \textbf{2} & 4 & 12 \\
        DNase (HL-60) & \textbf{2} & 3 & \textbf{2} & 11 \\
        DNase (K562) & 2 & 3 & \textbf{0} & 41 \\
        \bottomrule
    \end{tabular}\\
    Number of extraneous motifs (i.e., those which do not match any known relevant motif) found by each method for each dataset.
\end{table}

\begin{figure}[h]
\centering
\includegraphics[width=\columnwidth]{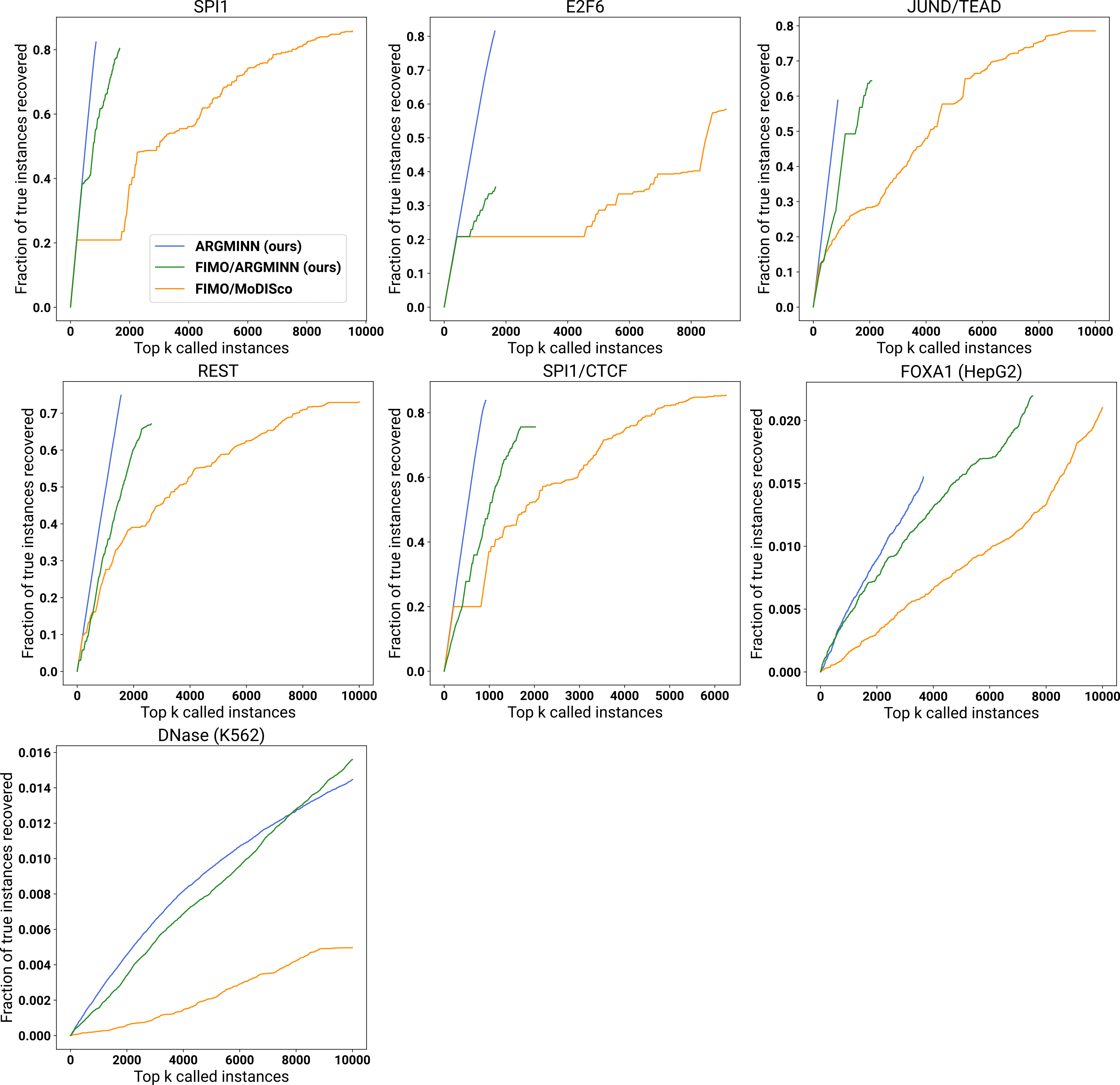}
\caption{\small We show additional examples of motif-instance quality, comparing ARGMINN with the traditional method of sequence-scanning using FIMO. As before, We rank motif instances by confidence (attention score from ARGMINN, or FIMO hit q-value), and compute the fraction of true motif instances that are covered in a top-$k$ fashion. We also compare our method to using FIMO to scan with ARGMINN-discovered motifs.}
\label{suppfig:more-motif-hits}
\end{figure}

\begin{table}[h]
    \caption{Motif-instance precision and recall}
    \label{supptab:motif-hit-pr}
    \centering
    \small
    \begin{tabular}{lcccccc}
        \toprule
        Dataset & \multicolumn{3}{c}{\textbf{Precision}} & \multicolumn{3}{c}{\textbf{Recall}} \\
        & A (ours) & F/AM (ours) & F/MM & A (ours) & F/AM (ours) & F/MM \\
        \midrule
        SPI1 & \textbf{0.953} & 0.881 & 0.852 & 0.824 & 0.804 & \textbf{0.858} \\
        TAL/GATA & \textbf{0.939} & 0.896 & 0.857 & \textbf{0.703} & 0.410 & 0.532 \\
        E2F6 & \textbf{0.954} & 0.804 & 0.777 & \textbf{0.816} & 0.355 & 0.584 \\
        JUND/TEAD & \textbf{1.000} & 0.920 & 0.980 & 0.588 & 0.644 & \textbf{0.785} \\
        REST & \textbf{0.972} & 0.853 & 0.891 & \textbf{0.749} & 0.671 & 0.731 \\
        SPI1/CTCF & \textbf{0.909} & 0.739 & 0.807 & 0.838 & 0.756 & \textbf{0.854} \\
        CTCF (HepG2) & 0.291 & 0.355 & \textbf{0.399} & \textbf{0.038} & 0.033 & 0.028 \\
        FOXA1 (HepG2) & 0.190 & \textbf{0.201} & 0.153 & 0.016 & \textbf{0.022} & 0.021 \\
        DNase (HepG2) & 0.229 & \textbf{0.386} & 0.196 & \textbf{0.014} & 0.012 & 0.001 \\
        DNase (K562) & 0.328 & \textbf{0.468} & 0.315 & 0.014 & \textbf{0.016} & 0.005 \\
        \bottomrule
    \end{tabular}\\
    Precision and recall values for motif instances discovered by ARGMINN (A) and by FIMO. We initialize FIMO either with motifs discovered by ARGMINN (F/AM), or with motifs discovered by MoDISco (F/MM).
\end{table}

\begin{figure}[h]
\centering
\includegraphics[width=\columnwidth]{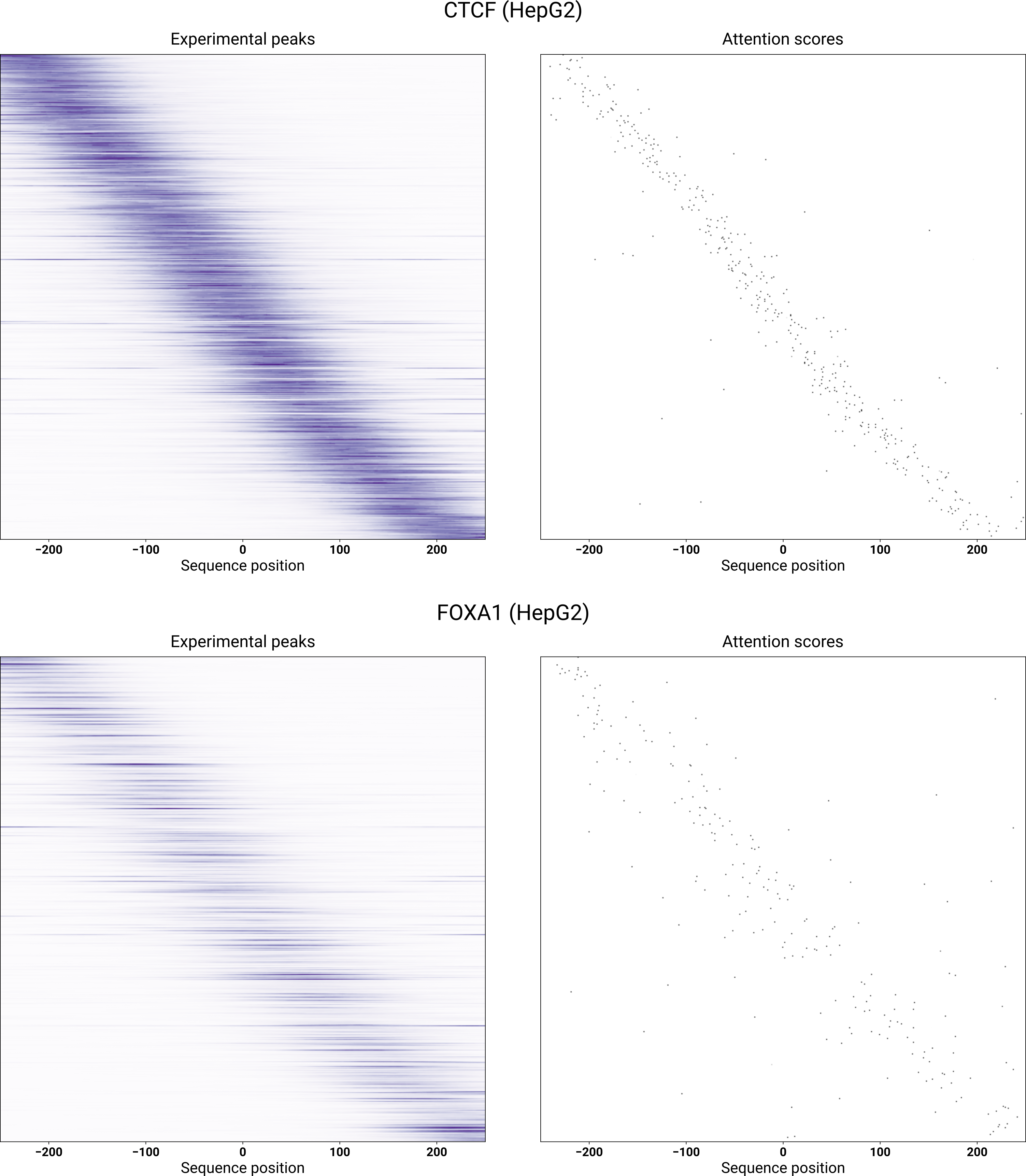}
\caption{\small We show the attention scores and the strength of the experimentally determined peaks at the test-set sequences for CTCF in HepG2 (above) and FOXA1 in HepG2 (below). In order to avoid center bias, the test peaks were independently and randomly jittered by up to 200 bp in either direction with uniform probability. The peaks/sequences are ordered left to right by the jittered peak summit. The attention scores closely track the peak locations across sequences, which demonstrates biological support for the interpretability of the attention scores.}
\label{suppfig:att-score-peaks}
\end{figure}

\begin{figure}[h]
\centering
\includegraphics[width=0.95\columnwidth]{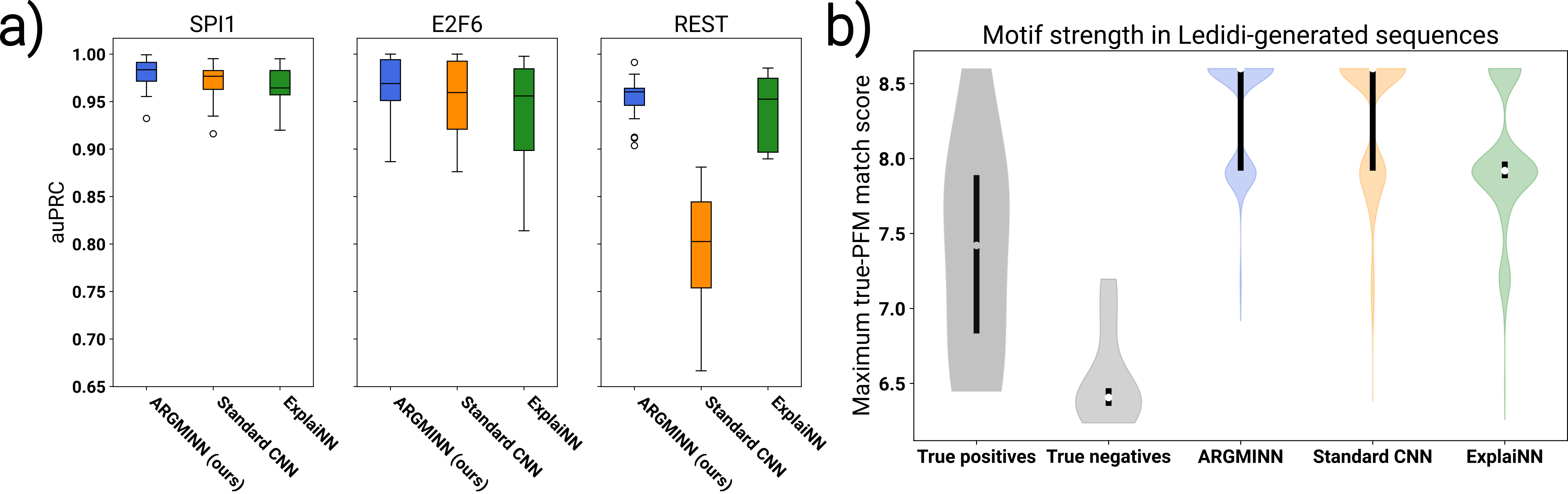}
\caption{\small \textbf{a)} After training on 50\% background GC content, we tested ARGMINN, a standard CNN, and ExplaiNN on varying levels of GC content, and show the resulting distribution of predictive performance. \textbf{b)} We used Ledidi to perform back-propagation-based sequence design on models trained to predict SPI1 binding, generating novel sequences which are meant to maximize the likelihood of SPI1 binding. We evaluated the quality of the generated sequences from each model by quantifying the distribution of match scores to the true SPI1 motif.}
\label{suppfig:robust}
\end{figure}

\begin{figure}[h]
\centering
\includegraphics[width=0.6\columnwidth]{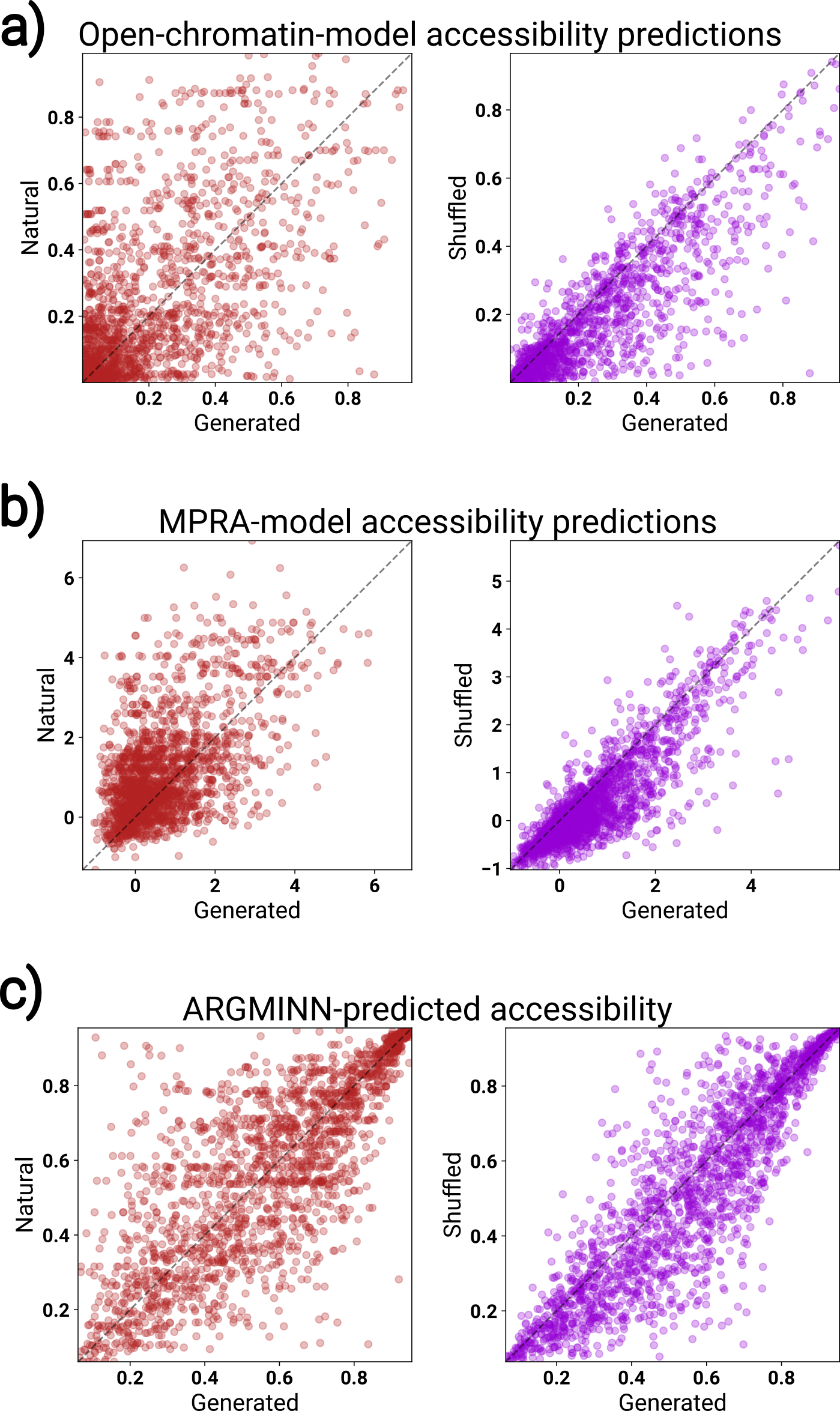}
\caption{\small We performed additional \textit{in silico} validation of our interpretably designed HepG2-accessible sequences. We tested our generated sequences using multiple independently trained models (trained on different datasets) as oracles: \textbf{a)} an open-chromatin model trained on an independently collected dataset of HepG2 accessibility; and \textbf{b)} an MPRA model trained on independently collected massively-parallel-reporter-assay data. We also show the predictions made by ARGMINN using the model as a predictive oracle rather than an interpretable-design agent (\textbf{c)}). In all cases, the generated sequences were far more accessible than background sequences, and were about as accessible as natural sequences identified by the experiment.}
\label{suppfig:seq-design-val}
\end{figure}

% \begin{figure}[h]
% \centering
% \includegraphics[width=\columnwidth]{figs/model_performance.png}
% \caption{\small We show the predictive performance of ARGMINN compared to the standard CNN and ExplaiNN. All models are given the same number and size of convolutional filters, and the overall complexity/capacity of the models are kept as similar as possible for comparison.}
% \label{suppfig:perf}
% \end{figure}

\FloatBarrier

\begin{table}[h]
    \caption{Model performance}
    \label{supptab:perf}
    \centering
    \small
    \begin{tabular}{lccccccccc}
        \toprule
        Dataset & \multicolumn{3}{c}{\textbf{Accuracy}} & \multicolumn{3}{c}{\textbf{auROC}} & \multicolumn{3}{c}{\textbf{auPRC}}\\
        \midrule
        % Dataset & ARGMINN & Standard CNN & ExplaiNN & ARGMINN & Standard CNN & ExplaiNN & ARGMINN & Standard CNN & ExplaiNN \\
         & A (ours) & SC & E & A (ours) & SC & E & A (ours) & SC & E \\
        SPI1 & \textbf{0.898} & 0.887 & 0.895 & \textbf{0.955} & 0.949 & 0.951 & \textbf{0.958} & 0.957 & 0.954 \\
        TAL/GATA & \textbf{0.884} & 0.838 & 0.856 & \textbf{0.954} & 0.910 & 0.939 & \textbf{0.959} & 0.918 & 0.941 \\
        E2F6 & \textbf{0.908} & 0.895 & 0.871 & \textbf{0.971} & 0.948 & 0.944 & \textbf{0.975} & 0.952 & 0.944 \\
        JUND/TEAD & \textbf{0.988} & 0.981 & 0.984 & \textbf{0.999} & 0.997 & 0.999 & \textbf{0.999} & 0.997 & 0.999 \\
        REST & \textbf{0.883} & 0.771 & 0.856 & \textbf{0.947} & 0.835 & 0.929 & \textbf{0.940} & 0.824 & 0.927 \\
        SPI1/CTCF & \textbf{0.909} & 0.860 & 0.900 & \textbf{0.968} & 0.931 & 0.955 & \textbf{0.965} & 0.928 & 0.953 \\
        CTCF (HepG2) & 0.769 & 0.775 & \textbf{0.782} & 0.848 & 0.850 & \textbf{0.859} & 0.854 & 0.852 & \textbf{0.860} \\
        FOXA1 (HepG2) & 0.728 & \textbf{0.734} & 0.730 & 0.805 & 0.809 & \textbf{0.810} & 0.798 & \textbf{0.806} & 0.804 \\
        DNase (HepG2) & 0.730 & 0.730 & \textbf{0.741} & 0.816 & 0.812 & \textbf{0.824} & 0.815 & 0.809 & \textbf{0.818} \\
        DNase (HL-60) & 0.753 & 0.760 & \textbf{0.769} & 0.833 & 0.842 & \textbf{0.853} & 0.826 & 0.839 & \textbf{0.847} \\
        DNase (K562) & 0.762 & 0.768 & \textbf{0.774} & 0.848 & 0.852 & \textbf{0.860} & 0.845 & 0.846 & \textbf{0.853} \\
        \bottomrule
    \end{tabular}\\
    Model performance comparing ARGMINN (A), a standard CNN (SC), and ExplaiNN (E). All models were given the same number and size of convolutional filters, and the overall complexity/capacity of the models were kept as similar as possible for comparison.
\end{table}

\begin{figure}[h]
\centering
\includegraphics[width=0.9\columnwidth]{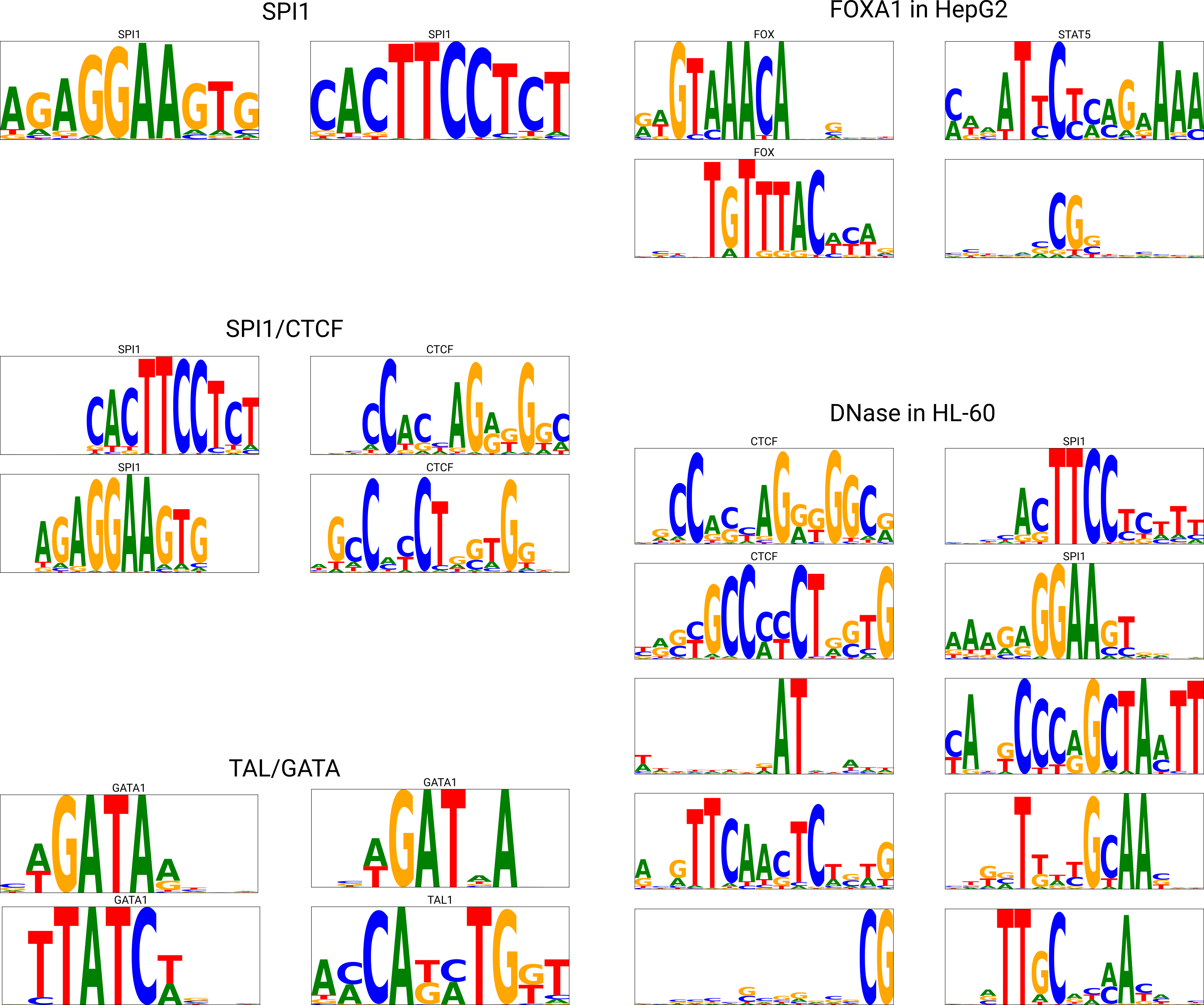}
\caption{\small We applied our filter regularizer (Equation \ref{eq:filterreg}, Equation \ref{eq:finalloss}) to the standard CNN architecture, and show the resulting motifs extracted from the first-layer filters. With the filter regularization, the standard CNN attained the ability to show clean, non-redundant, relevant motifs in its filters. Thus, our filter regularizer is able to turn even traditional neural networks into more mechanistically interpretable architectures. Importantly, however, although a standard CNN with our filter regularizer can now reveal discovered motifs, without our unique attention mechanism (Equation \ref{eq:att}), it still is unable to easily reveal motif instances and syntax. As such, these partially interpretable architectures would still need to rely on traditional motif-instance-scanning algorithms.}
\label{suppfig:cnn-filterreg}
\end{figure}

\begin{figure}[h]
\centering
\includegraphics[width=\columnwidth]{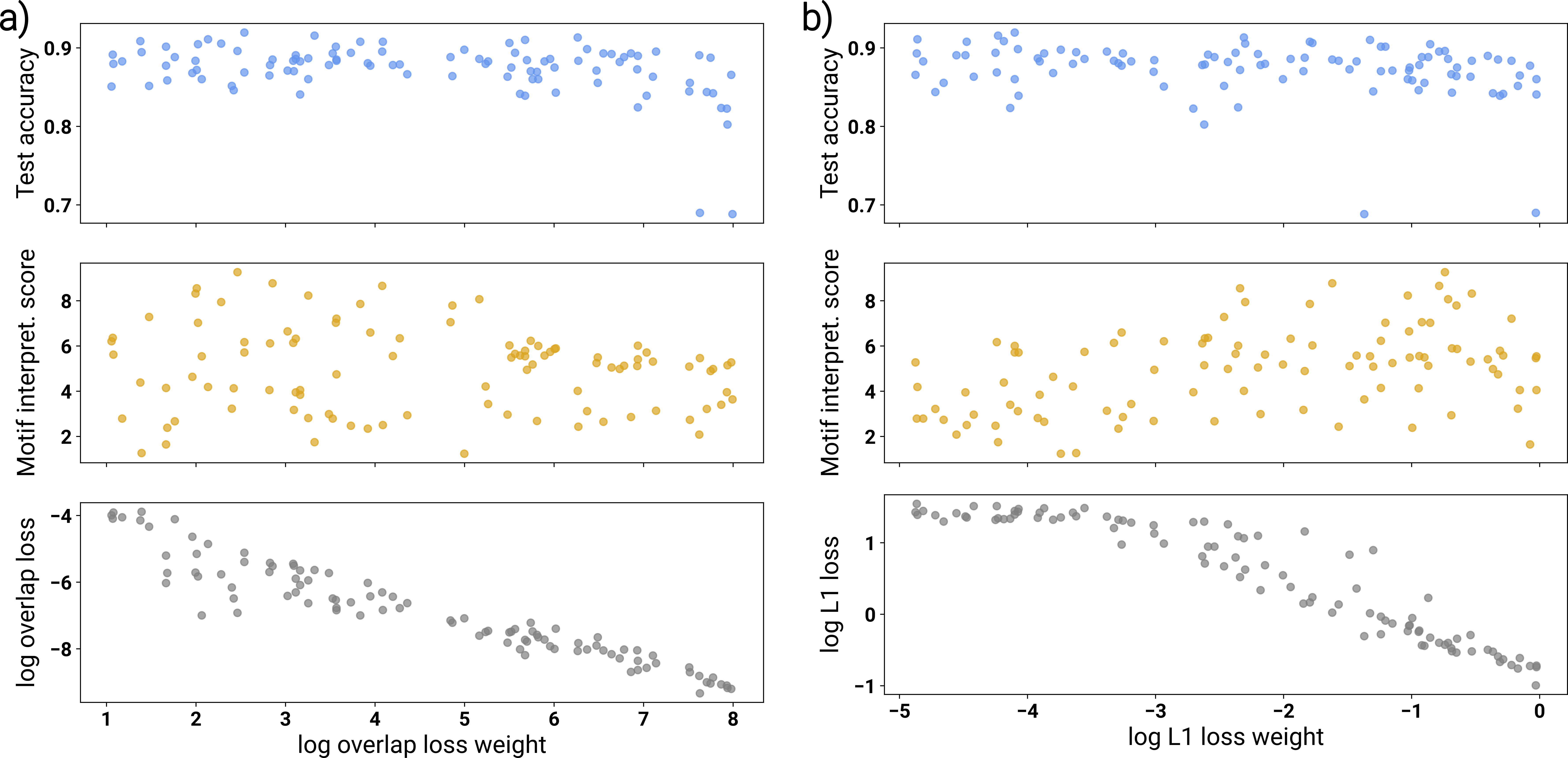}
\caption{\small We show the effect of the filter-overlap (left) and filter-L1 loss (right) weights on the performance and interpretability of ARGMINN. We also show the effect of the loss weights on the value of the filter-overlap and filter-L1 loss itself. Interpretability is measured by the maximum similarity of the discovered motifs to the true motif. In general, ARGMINN’s predictive performance and interpretability remained largely invariant to the loss weights over a wide range of orders of magnitude.}
\label{suppfig:loss-weight-robust}
\end{figure}

\newpage
\clearpage

\section{Supplementary Methods}
\label{app:supp-methods}

All of the code used to generate the results and figures in this paper is available here:

\texttt{https://github.com/Genentech/ARGMINN}

\subsection{Training data}
\label{app:supp-methods:data}

For our simulated datasets, we downloaded motif PFMs (position frequency matrices) from JASPAR \citep{Fornes2020}, and trimmed off low-information-content flanks. When training and testing, we randomly generated 500 bp sequences on the fly. Motif instances were sampled from the PFMs, and inserted into a randomly sampled background (from a uniform distribution of A, C, G, and T). Motifs (or combinations of them) were randomly inserted in the central 100 bp of the background. Our simulations contained motif configurations as follows:
\begin{itemize}
    \item SPI1: all positive sequences have a single instance of the SPI1 motif
    \item TAL/GATA: 37.5\% of positive sequences have a single instance of TAL1, 37.5\% have a single instance of GATA1, and 25\% of sequences have both (either 7, 8, or 9 bp apart)
    \item E2F6: 10\% of positive sequences have a single instance of E2F6, and 90\% have both E2F6 and MAX, between 30 and 60 bp apart; 50\% of negative sequences have only the MAX motif, and the other 50\% are random background
    \item JUND: 25\% of positive sequences have the JUND TRE motif, 25\% have the JUND CRE motif, and 50\% have the JUND TRE motif followed by the TEAD4 motif with a spacing of 6 bp in between
    \item REST: all positive sequences have both the left- and right-half motifs, with a spacing of 2, 6, 7, 8, 9, 10, or 11 bp apart; 25\% of negative sequences have only the left-half motif, 25\% have only the right-half motif, and 50\% are random background
    \item SPI1/CTCF: all positive sequences have a single instance of the SPI1 motif; 50\% of negative sequences have SPI1 followed by the CTCF motif (either 30, 40, or 50 bp apart), and 50\% are random background
\end{itemize}

When generating simulated sequences, we scanned the random backgrounds for spurious matches and filtered out such instances. Note that this is a crucial step for simulated datasets, as for short motifs such as GATA (6 bp), we would expect over 12\% of random backgrounds to contain a perfect match by chance. This is an issue which is much rarer in real datasets, but for our simulations, we scanned the PFMs (and their reverse complements) across our generated backgrounds (for positive and negative sequences), and ensured that no PFM which was used in the simulation attained a match score of over 0.9.

For our experimental datasets, we downloaded the IDR peaks from ENCODE \citep{ENCODE_Project_Consortium2012-wb} for the following experiments:

\begin{table}[h]
    \caption{Experimental ENCODE datasets}
    \centering
    \small
    \begin{tabular}{lll}
        \toprule
        Dataset & ENCODE experiment ID & IDR peaks file ID \\
        \midrule
        CTCF (HepG2) & ENCSR607XFI & ENCFF664UGR \\
        FOXA2 (HepG2) & ENCSR865RXA & ENCFF081USG \\
        REST (K562) & ENCSR054JMQ & ENCFF118ECK \\
        DNase (HepG2) & ENCSR149XIL & ENCFF897NME \\
        DNase (HL-60) & ENCSR889WKL & ENCFF773SFA \\
        DNase (K562) & ENCSR000EKS & ENCFF274YGF \\
        DNase (GM12878) & ENCSR000EMT & ENCFF073ORT \\
        \bottomrule
    \end{tabular}\\
\end{table}

Our positive dataset consisted of random 500 bp sequences drawn from the genome, where at least half of the 500 bp overlaps a peak (if the peak is less than 500 bp), or at least half of the peak overlaps the 500 bp (if the peak is over 500 bp). Our negative dataset consisted of randomly sampled intervals from the genome. If a randomly selected negative sample overlapped a peak by more than half (or \textit{vice versa}, it was relabeled as positive for the batch).

Note that when training, we automatically used reverse-complement augmentation so that every batch contained sequences along with their reverse complements.

\subsection{Model architectures}
\label{app:supp-methods:arch}

Our ARGMINN architecture consists of two modules: motif scanners and a syntax builder. The motif scanners consist of a single convolutional layer of $n_{f}$ filters, each of $w$ bp in width (typical values are $n_{f} = 8$, $w = 10$), which scan across a one-hot-encoded DNA sequence. The result is passed to a ReLU, and the output constitutes the ``motif activations''.

The motif activations are concatenated with a positional encoding of dimension $d = 16$. Our positional encoding is defined as follows:

$$P_{i,2j} = \sin(\frac{i}{10000^{\frac{2j}{d}}}), P_{i,2j+1} = \cos(\frac{i}{10000^{\frac{2j}{d}}})$$

where $i$ is the position along the sequence, and $j \in \{0,\ldots,\frac{d}{2} - 1\}$.

Let $A\Vert P$ be the motif activations and positional encodings concatenated together. This is passed to two consecutive memory-stream-based attention layers. The $l$th attention layer starts with a memory stream $m_{l-1}$ of dimension 128 ($m_{0}$ is a vector of all 1s), and $A\Vert P$. In each layer, $m_{l-1}$ is passed through a linear layer to obtain a single query $q_{l}$ of dimension equal to the dimension of each input token in $A\Vert P$. Two separate linear layers also convert each token in $A\Vert P$ into a set of key vectors and value vectors (of the same dimension as the input token). The query, keys, and values are reshaped to obtain 4 attention heads. We then compute the attention scores by multiplying the query against all keys, and normalizing by $\sqrt{d_{q_{l}}}$, where $d_{q_{l}}$ is the dimension of the query/key/value vectors. We then perform dropout on the attention scores with dropout rate 0.1, and softmax the scores for each attention head. These attention scores are used to weight the value vectors in a weighted sum, which is then reshaped to reincorporate the heads. This is passed through another linear layer which retains the same dimension, followed by dropout and layer normalization. This is then fed to a 2-layer MLP with ReLU and dropout in between the two linear layers, mapping the result to the same dimension as $m_{l-1}$. After a final dropout and layer norm, this is added to $m_{l-1}$ to obtain $m_{l}$.

After all attention layers, the final $m_{l}$ is passed to a single linear layer which maps it to a scalar prediction which is passed to a sigmoid activation function (for binary prediction).

Our standard CNN follows an architecture which is common in the literature for single-task predictions. We apply 3 successive convolutional layers to the input sequence, each with $n_{f}$ filters. The filters of the first layer have width $w$. The next two convolutional layers have filters of width 5. After each layer we apply ReLU and batch normalization. We then perform max pooling with a filter size of 40 and a stride of 40. This is passed to two linear layers of 10 and 5 hidden dimensions each. After each linear layer, we apply ReLU and batch normalization. Finally, a final linear layer maps the result to a sigmoid-activated output.

Our ExplaiNN implementation follows the description in \citet{Novakovsky2022}. As with the other architectures, we use the same number and width of first-layer convolutional filters.

Depending on the complexity of the dataset's syntax, we selected the ARGMINN architectural hyperparameters accordingly. Note that these values were not tuned at all, and were chosen at the outset based on domain knowledge and never modified. We also ensured that for each dataset, we always used the same number and length of first-layer convolutional filters ($n_{f}$ and $w$, respectively) in other architectures for a fair comparison.

\begin{table}[h]
    \caption{ARGMINN architectural hyperparameters}
    \centering
    \small
    \begin{tabular}{lccc}
        \toprule
        Dataset & Number of attention layers $n_{L}$ & Number of filters $n_{f}$ & Length of filters $w$ \\
        \midrule
        SPI1 & 1 & 8 & 10 \\
        TAL/GATA & 1 & 8 & 10 \\
        E2F6 & 2 & 8 & 10 \\
        JUND & 1 & 8 & 10 \\
        REST & 2 & 8 & 10 \\
        SPI1/CTCF & 2 & 8 & 15 \\
        CTCF (HepG2) & 1 & 8 & 15 \\
        FOXA2 (HepG2) & 2 & 8 & 15 \\
        REST (K562) & 2 & 8 & 8 \\
        DNase (HepG2) & 3 & 20 & 15 \\
        DNase (HL-60) & 3 & 20 & 15 \\
        DNase (K562) & 3 & 20 & 15 \\
        DNase (GM12878) & 3 & 20 & 15 \\
        \bottomrule
    \end{tabular}\\
\end{table}

\FloatBarrier

\subsection{Training schedules}
\label{app:supp-methods:training}

We trained all of our models and performed all analyses on a single Nvidia A100.

When training, we used a batch size of 128 with an equal number of positives and negatives (this includes the reverse-complement augmentation). For our simulated datasets, each training epoch consisted of 100 batches, and each validation and test epoch consisted of 10 batches. For our experimental datasets, we reserved chr8 and chr10 for validation, and chr1 for test (all other autosomes along with chrX were used for training).

We trained all of our models for 40 epochs (regardless of architecture), and noted that the loss had converged in all cases. We used a learning rate of 0.001.

For ARGMINN, we weighted the secondary losses as follows:

\begin{itemize}
    \item $\lambda_{o}$: 0 for the first 10 epochs, increasing from $10^{0.5}$ to $10^{4}$ evenly in logarithmic space over the next 20 epochs, and stable at $10^{4}$ for the last 10 epochs
    \item $\lambda_{l}$: 0 for the first 10 epochs, increasing from $10^{-4}$ to $10^{-3}$ evenly in logarithmic space over the next 20 epochs, and stable at $10^{-3}$ for the last 10 epochs
\end{itemize}

For all of our models (ARGMINN, standard CNN, ExplaiNN), we trained 3 random initializations and selected the one with the best test accuracy for downstream analyses.

\subsection{Analyses}
\label{app:supp-methods:analyses}

\textbf{Extracting motifs from convolutional filters}

To extract motif PFMs from convolutional filters, we adopted a procedure based on that described in \citet{Kelley2016}. Specifically, we passed the test set through the model and computed the average of all sub-sequences (pooling together all possible sub-sequences in the test set) which activated that filter to at least 50\% of the maximum activation achieved over all such windows.

\textbf{Computing importance scores}

We computed importance scores using DeepLIFTShap, integrated gradients, or \textit{in silico} mutagenesis \citep{Shrikumar2017,Sundararajan2017}.

For DeepLIFTShap and integrated gradients, we used PyTorch Captum. For DeepLIFTShap, we used a reference of 10 baselines consisting of dinucleotide-shuffled sequences, as recommended in \citet{Shrikumar2017}. We also recovered the hypothetical importance scores at each position, as recommended for MoDISco \citep{Shrikumar2018}. For integrated gradients, we used a baseline of all 0s.

To obtain \textit{in silico} mutagenesis scores, we computed the importance of a base by first computing the difference between the output prediction of the original sequence versus every possible mutation that made at that position. We mean normalized over the base dimension to obtain a set of hypothetical importance scores. The actual importance score for a position was simply the hypothetical importance score (after mean normalization) for the base actually present in the sequence at the position.

\textbf{Discovering motifs with MoDISco}

In general, we discovered motifs using MoDISco from the standard CNN.

We computed importance scores over the entire positive-labeled dataset using the DeepLIFTShap algorithm as described above. We then ran MoDISco-lite v2.2.0 \citep{Shrikumar2018} using a maximum of 10000 seqlets and default parameters, as recommended by the authors.

\textbf{Evaluating discovered motifs}

To compute matches to known motifs, we used TOMTOM \citep{Bailey2015} to compute the q-value similarity between a PFM to known motifs (across all possible alignments). We reported the $-\log_{10}(q)$ value as similarity. For simulated datasets, we reported the closest match (i.e., highest similarity) to any motif in the dataset. For experimental datasets, we reported the closest match to any relevant motif in the JASPAR human-motif database \citep{Fornes2020}.

Note that to ensure a fair comparison, the lengths of the motifs were kept the same in each dataset. For ARGMINN, the standard CNN, and ExplaiNN, all models are trained with the same first-layer filter sizes. MoDISco by default outputs longer patterns, so we trimmed MoDISco-discovered motifs to the same size as the filters used by other methods (maximizing the total information content in the post-trimmed window; we used a uniform background for computing information content).

For our analyses on motif accuracy/similarity, redundancy, and number of extraneous motifs, we needed to match each motif discovered by each method to the closest \textit{relevant} motif (or none at all). To do this, we first needed to identify the set of possible relevant motifs for each dataset (i.e., the set of motifs which would be considered biologically ``accurate'' for the task). For simulated datasets, the set of relevant motifs was simply the PFMs used to create the simulation. For experimental datasets, the set of relevant motifs was defined by first running TOMTOM against all known human motifs and pooling together the top matches (by motif family) over all methods and architectures (e.g., ARGMINN, MoDISco, etc.). We used a q-value cut off of 0.5. After pooling together all the top TOMTOM matches over all methods, we extracted a set of relevant motifs or motif families by checking for supporting literature. Any motif/family with supporting literature was kept as a relevant motif.

Finally, once the set of relevant motifs for each dataset was determined, we matched each motif (discovered by each method) to the closest relevant motif using TOMTOM. Discovered motifs which did not match any relevant motif (using the default TOMTOM threshold) were considered extraneous.

To compute redundancy, we counted the number of times each relevant motif was matched to by a method's discovered motifs. We kept track of forward and reverse-complement orientations. As long as one orientation was discovered, we considered that motif to had been found; we computed redundancy as the maximum number of times a relevant motif was matched to (maximum over orientations). For reverse-complement symmetric motifs/families, we did not consider orientations separately, and computed redundancy accordingly.

To compute motif accuracy, we computed the similarity (measured by the TOMTOM q-value) of the closest motif discovered by each method to each relevant motif.

\textbf{Tracing back motif instances and syntax}

To trace back motif instances for a particular input sequence in ARGMINN, we performed a forward pass and retained the motif activations and attention scores. Over all layers and all attention heads, we examined the positions in the sequence which had an attention score of at least 0.9. We then called a motif hit if the activation for a filter at that position was at least the average activation (for that filter) over the test set.

In our analyses, we called motif instances over the test set.

To identify syntax, we separated the input sequences by which motif instances were called, and computed the distribution of the spacings between the motif instances.

In order to rank motif instances from ARGMINN, we ranked by maxmimum attention score over all heads/layers at that position. To break any ties, we used the highest motif-filter activation score at that position.

\textbf{Scanning for motif instances with FIMO}

Before running FIMO on MoDISco or ARGMINN motifs, we trimmed and filtered the motifs for high-information-content regions. Specifically, we cut off flanks with information content lower than 0.2. We then required that after trimming, the motif was at least 5 bp and had an average information content of 1.0. We used a uniform background for computing information content.

To scan for motif instances using FIMO, we started with PFMs and ran FIMO v5.0.5 \citep{Bailey2015} on test-set sequences and their reverse complements. We used the default FIMO parameters.

To rank FIMO hits, we used the q-value from FIMO. We also collapsed overlapping FIMO hits before analyzing, keeping the most significant q-value between overlapping hits for ranking purposes.

\textbf{Evaluating motif instances}

To evaluate our motif instances, we compared called motif instances to ground-truth instances. We computed the precision as the fraction of called instances which overlap ground-truth instances, and the recall as the fraction of ground-truth instances which overlap called instances. We also computed recall curves in a top-$k$ fashion, where we ranked the called instances (described above) and for each top $k$ called instances, we computed the recall relative to ground-truth instances.

For simulated datasets, the ground-truth motif instances are completely known, as they are defined at the time of sequence generation.

For experimental datasets, we obtained ``ground-truth'' motif instances from \citet{Vierstra2020}, an independently collected set of DNA-binding footprints in various cell types. For each cell type of interest (e.g., HepG2), we simply pooled together the footprints of all experiments from that cell type and used those as a set of ground-truth motif instances.

Due to computational efficiency, we also limited the set of called motif instances for experimental datasets to the top 10000 hits.

\textbf{Evaluating QTL prioritization}

We downloaded causal and non-causal QTLs in the GM12878 cell type from \citet{Lee2015}. We limited the set of dsQTLs to only those in the test set of our models (i.e., on chr1).  We then took our models trained on GM12878 DNase accessibility, and for each putative QTL (causal or non-causal), we computed the absolute difference in the output prediction, and treated that as a score to compute precision and auROC.

\textbf{Computing GC-content robustness}

To test the robustness of our models against changes in GC content, we examined our models trained on simulated datasets (which allow us to modify the background GC content freely while keeping the motifs the same). Without retraining or fine-tuning, we evaluated the models' predictive performance on background GC content of 5\%, 10\%, 15\%, \ldots, 90\%, 95\%.

\textbf{Generating sequences using Ledidi}

To generate novel sequences using Ledidi, we took our models (ARGMINN, standard CNN, ExplaiNN) trained on our simulated SPI1 dataset. For each model, we fitted Ledidi 32 times, and for each fitted instance, we generated 32 sequences. This yielded 1024 Ledidi-generated SPI1 sequences for each model. We used $\tau = 5, \lambda = 5000$ for Ledidi.

To evaluate the quality of the generated sequences, we scanned the true PFM across each sequence (note that these models were all trained on sequences which contain motifs sampled from this exact PFM), and computed the top match score (as a cross-correlation score) for each sequence.

We computed significance of the difference in distributions (between different models used with Ledidi) of top SPI1 match scores using a one-sided Mann-Whitney U test.

\textbf{Generating adversarial examples}

We generated adversarial examples in a standard CNN in two ways.

First, we generated sequences which had no motifs present, but were still predicted to have a positive label by the CNN. To do this, we built up random sequences by taking highly-activating sequences from random filters and adding them to the sequence. We took the most highly-activating sequence for each filter and randomly strung together a random ordering of such sequences to obtain a 500 bp example. We verified that no part of these sequences matched any motif (or reverse complement) in the dataset, using the same criteria as the data loader as described above.

Next, we generated sequences which had motifs, but were still predicted to be negative by the CNN. To do this, we first sampled a motif configuration from the normal simulation. Under normal circumstances, such a configuration would endow a positive label in the dataset. We then surrounded either side of this configuration with sequences that are least activating for a random ordering of filters. We randomly strung together a random ordering of such sequences to pad out a 500 bp example.

\textbf{Interpretable sequence design}

To design novel sequences intrepretably using ARGMINN, we started with our ARGMINN model trained on HepG2 DNase accessibility. We extracted motifs from the filters as described above. Prior to designing, we trimmed and filtered the motifs. Specifically, we first cut off flanks with information content lower than 0.2. We then required that after trimming, the motif was at least 5 bp and had an average information content of 0.8. We used a uniform background for computing information content.

We then extracted the motif syntax as described above, again only using the test set. This procedure takes each input sequence and assigns it a motif pattern (e.g., ``CTCF - TEAD'' or ``FOXA'' or ``CTCF - FOXA - FOXA''). The motif pattern denotes which motifs were identified by ARGMINN in that sequence (in order) (e.g., CTCF followed by TEAD). Over the entire test set, we labeled each sequence with its motif pattern. For each pattern which has more than one motif, ARGMINN then gives a set of spacings that were identified over all sequences with that pattern.

We took the top 20 patterns (by number of sequences which have that pattern, requiring at least 5 sequences with that pattern). For each of the top 20 patterns, we generated 100 novel sequences with that pattern. To construct a novel sequence for a pattern, we started with an original 500 bp test sequence which falls under that pattern (sampled randomly). We then dinucleotide-shuffled the central 300 bp to destroy any functional motifs. We then inserted the motifs of that pattern (sampling from the ARGMINN-learned PFM(s)), at spacings sampled from the empirical distribution for that pattern (for single-motif patterns, the motif was simply placed in the center).

To validate our generated sequences \textit{in silico}, we used several independently trained models as oracles:

\begin{enumerate}
    \item Interpretations from ARGMINN itself were used to generate these sequences, but we also used ARGMINN as a predictive oracle, as it predicts accessibility from DNA sequence.
    \item We used Borzoi \citep{Linder2023} as a predictive oracle. Specifically, we used output task 1510 (trained on ENCFF577SOF), which also measured HepG2 DNase accessibility. Whereas we designed 500 bp sequences, Borzoi predicts entire signal-profile tracks from an extremely large context (524000 bp inputs). To evaluate our 500 bp sequences, we took the original coordinate from which the sequence arose, and padded it with the appropriate genomic context on either side to from the full 524 kb input. To produce a scalar prediction, we summed the center of the (binned) profile prediction corresponding to the central 500 bp input.
    \item We fine-tuned Enformer \citep{Avsec2021} on HepG2 DNA accessibility measured by massively parallel reporter assays (MPRAs). We downloaded the MPRA data from \citet{Gosai2023}, and verified that the predictive performance was on-par with the performance recorded in \citet{Gosai2023}. The fine-tuning was performed using the gReLU software package.
    \item We fine-tuned Enformer using gReLU, using binary labels of HepG2 DNase accessibility from a different ENCODE expriment (ENCSR291GJU) from the one ARGMINN was trained on.
\end{enumerate}

We computed significance of the difference in predicted accessibility (e.g. generated vs. background, or generated vs. natural) using a one-sided Wilcoxon test.

\textbf{Expressivity requirement for REST binding}

In order to demonstrate the expressivity and interpretability of ARGMINN compared to ExplaiNN, we constructed a simulated REST dataset which explicitly tests the requirements of the unique binding syntax of REST. In this dataset, positive sequences always had the left and right motifs (in that order) with 10 bp in between. Negative sequences were structured as follows: 10\% have only the left motif; 10\% have only the right motif; 50\% have both the left and right motifs but in improper order or orientation (e.g., the reverse-complement of left and right in that order, or the left with the reverse of the right, etc.), with each configuration having a 10 bp spacer and being selected uniformly at random; 30\% have two left or two right motifs, of all sorts of configurations (e.g., forward or reverse), each having a 10 bp spacer and being selected uniformly at random; 10\% are uniform background.

\textbf{Filter regularization on standard CNN}

We applied our filter regularization to a standard CNN architecture, using the same loss weights and annealing schedule as described above. We assigned identities to these motifs also using the procedure with TOMTOM as described above.

\textbf{ARGMINN loss-weight robustness}

To evaluate the robustness of ARGMINN to the weights of the filter-overlap loss and the filter-L1 loss, we trained ARGMINN on the simulated SPI1 dataset 100 times, each time with a randomly selected loss weight. The filter-overlap loss weight was randomly selected between $\lambda_{o}\in [10^{1}, 10^{8}]$ and the filter-L1 loss weight was randomly selected between $\lambda_{l}\in[10^{-5}, 10^{0}]$, where sampling was done uniformly on a logarithmic scale. We trained each model to completion as described above.

To evaluate the motif interpretability of each model, we extracted motifs from each ARGMINN model as described above. For each model, we then we computed the average similarity of all the discovered motifs to the true SPI1 PFM (similarity was computed as the $-\log(q)$ score from TOMTOM). If a discovered motif was not similar enough to the true PFM to pass TOMTOM's thresholds, it was given a $-\log(q)$ value of 0 by default.

\end{document}